
\input harvmac

\def\perpp{{\scriptscriptstyle\perp}}

\def\half{{1\over 2}}

\def\kbT{k_{\scriptscriptstyle\rm B}T}

\def\bo#1{{\cal O}(#1)}

\def\free{\hbox{$\cal F$}}
\def\bold#1{\setbox0=\hbox{$#1$}%
     \kern-.010em\copy0\kern-\wd0
     \kern.025em\copy0\kern-\wd0
     \kern-.020em\raise.0200em\box0 }
\def\nabb{\bold{\nabla}}
\def\dz{\partial_z}

\def\dx{\partial_x}
\def\dy{\partial_y}
\def\di{\partial_i}

\def\kbt{k_{\rm B}T}
\def\ts{\theta_6}
\def\dn{\delta n}
\def\angstrom{{\hbox{\AA}}}
\def\grad{\vec\nabla_{\!\!\perp}}
\def\cross{\!\!\times\!\!}
\def\dnb{\delta {\vec n}}
\def\dot{\!\cdot\!}

\def\curl{\nabb\cross}
\lref\BOL{
W.A.~Bollman, {\sl Crystal Defects and Crystalline Interfaces}
(Springer-Verlag,
Berlin, 1970);
J.P.~Hirth and J.~Lothe, {\sl Theory of Dislocations}, Second ed. (Wiley, New
York,
1982).
}
\lref\KLEMi{M. Kl\'eman, J. Phys. (Paris) {\bf 46}, 1193 (1985).}
\lref\SWM{S.~Meiboom, J.P.~Sethna, P.W.~Anderson and W.F.~Brinkman, Phys. Rev.
Lett. {\bf 46},
1216 (1981);
D.C.~Wright and N.D.~Mermin, Rev. Mod. Phys. {\bf 61}, 385 (1989).}
\lref\MN{M.C.~Marchetti and D.R.~Nelson, Phys. Rev. B {\bf 41}, 1910 (1990).}
\lref\NELii{D.R.~Nelson and L.~Peliti,  J. Phys. (Paris) {\bf 48}, 1085 (1987);
see also F.~David, {\sl Statistical Mechanics of Membranes and Surfaces :
Jerusalem Winter School for Theoretical Physics}, edited by D.R.~Nelson,
{\sl et. al.} (World Scientific, Singapore, 1989).}
\lref\TGB{S.R.~Renn and T.C.~Lubensky, Phys. Rev. A {\bf 38}, 2132 (1988);
{\bf 41}, 4392 (1990).}
\lref\KN{R.D.~Kamien and D.R.~Nelson, Phys. Rev. Lett. {\bf 74}, 2499 (1995).}
\lref\KNW{D.R.~Nelson
and R.D.~Kamien, to appear in {\sl Proceedings of the Wiener 1994 Centennial
Symposium},
edited by D.~Jerison,
I.M.~Singer and D.W.~Strock (American Mathematical Society, Providence,
1995).}
\lref\TLL{R.D.~Kamien and T.C.~Lubensky, J. Phys. I (Paris) {\bf 3}, 2131
(1993).}
\lref\DRN{D.R.~Nelson, Phys. Rev. Lett. {\bf60}, 1973 (1988);
D.R.~Nelson and H.S. Seung, Phys. Rev. B {\bf 39}, 9153 (1989).}
\lref\KLN{P. Le~Doussal and D.R. Nelson, Europhys. Lett. {\bf 15}, 161 (1991);
R.D. Kamien, P. Le~Doussal, and D.R.~Nelson, Phys. Rev. A {\bf
45}, 8727 (1992); Phys. Rev. E {\bf 48}
,4116 (1993).}
\lref\TON{J.~Toner, Phys. Rev. A {\bf 27}, 1157 (1983).}
\lref\TONP{We thank J.~Toner for discussions on this point.
}
\lref\DGP{P.G.~de Gennes and J.~Prost, {\sl The Physics of Liquid Crystals},
Second
Edition,
Chap. VII
(Oxford University Press, New York, 1993).}
\lref\DGS{P.G.~de Gennes, Solid State Commun. {\bf 14}, 997 (1973).}
\lref\BP{A.~Le~Forestier and F.~Livolant, Liq. Cryst. {\bf 17}, 651 (1994).}
\lref\PR{J.~Prost, Liq. Cryst. {\bf 8}, 123 (1990).}
\lref\KLEM{M.~Kl\'eman, Rep. Prog. Phys. {\bf 52}, 555 (1989).}
\lref\KO{M.~Kl\'eman and P.~Oswald, J. Phys. (Paris) {\bf 43}, 655 (1982).}
\lref\BOUi{Y.~Bouligand, J. Phys. (Paris) {\bf 41}, 1297 (1980).}
\lref\BOUii{F.~Livolant
and Y.~Bouligand, J. Phys. (Paris) {\bf 47}, 1813 (1986); F.~Livolant,
J.~Mol.~Biol.
{\bf 218} 165 (1991).}
\lref\PARS{R.~Podgornik and V.A.~Parsegian, Macromolecules {\bf 23}, 2265
(1990).}
\lref\GIA{C.~Gianessi, Phys. Rev. A {\bf 28}, 350 (1983); Phys. Rev. A {\bf
34},
705 (1986).}
\lref\NT{D.R.~Nelson and J.~Toner, Phys. Rev. B {\bf 24}, 363 (1981).}
\lref\IND{V.L.~Indenbom and A.N.~Orlov, Usp. Fiz. Nauk {\bf 76}, 557 (1962) [
Sov. Phys. Uspekhi {\bf 5}, 272 (1962)].}
\lref\KOS{A.M.~Kosevich, Usp. Fiz. Nauk {\bf 84}, 579 (1964) [Sov. Phys.
Uspekhi
{\bf 7},
837 (1965)].}
\lref\MH{N.D.~Mermin and T.L.~Ho, Phys. Rev. Lett. {\bf 36}, 594 (1976).}
\lref\SB{J.V.~Selinger and R.F.~Bruinsma, Phys. Rev. A {\bf 43}, 2910 (1991).}
\lref\HN{See, for example \NT and B.I.~Halperin, unpublished.}
\lref\VOL{G.E.~Volovik, {\sl Exotic Properties of $^3$He}, Chap. III
(World Scientific, Singapore, 1992).}
\lref\PP{G.A.~Hinshaw, Jr., R.G.~Petschek and R.A.~Pelcovits, Phys. Rev. Lett.
{\bf 60},
1864 (1988).}
\lref\AL{B.~Alberts, {\sl et. al.}, {\sl Molecular Biology of the Cell}
(Garland, New York, 1989).}
\lref\SEL{J.V.~Selinger and J.M.~Schnur, Phys. Rev. Lett. {\bf 71}, 4091
(1993); J.V. Selinger, Z.-G.~Wang, R.F.~Bruinsma and C.M.~Knobler,
Phys. Rev. Lett. {\bf 70}, 1139 (1993).}
\lref\PN{P.~Nelson and T.~Powers, Phys. Rev. Lett. {\bf 69}, 3409 (1992);
J. Phys. II (Paris) {\bf 3}, 1535 (1993).}
\lref\HPP{G.A.~Hinshaw, R.G.~Petschek and R.A.~Pelcovits,
Phys. Rev. Lett. {\bf 60}, 1864 (1988).}
\lref\SL{S.~Langer and J.~Sethna, Phys. Rev. A {\bf 34},
5035 (1986).}
\lref\RSDV{R.L.~Rill, T.E.~Strzelecka, D.H.~Van Winkle and M.W.~Davidson,
Physica A, {\bf 176}, 87 (1991).}
\lref\MEYER{R.B.~Meyer, {\sl Polymer Liquid
Crystals}, edited by A. Ciferri, W.R. Kringbaum and R.B. Meyer (Academic, New
York, 1982) Chapter 6.}
\lref\TER{E.M.~Terentjev, Europhys. Lett. {\bf 23}, 27 (1993).}
\lref\KNT{R.D.~Kamien, Institute for Advanced Study Preprint, IASSNS-HEP-95/44,
(1995) [cond-mat/9507023].}
\lref\TM{V.G.~Taratura and R.B.~Meyer, Liquid Crystals {\bf 2}, 373
(1987).}
\lref\TONER{J.~Toner, Phys. Rev. Lett. {\bf 68}, 1331 (1992).}
\lref\RLRAT{T.C.~Lubensky, T.~Tokihiro and S.R.~Renn, Phys. Rev. Lett. {\bf
67}, 89
(1991).}
\lref\KT{R.D.~Kamien and J.~Toner, Phys. Rev. Lett. {\bf 74} 3181 (1995).}
\lref\Rix{H.~Block,
{\sl Poly($\gamma$-Benzyl-L-Glutamate) and other Glutamate Acid Containing
Polymers} (Gordon and Breach, London, 1983).}
\lref\Rxi{R.B.~Meyer, F.~Lonberg,
V.~Tarututa, S.~Fraden, S.D.~Lee and
A.J.~Hurd, Disc. Faraday Chem. Soc. {\bf 79} 125 (1985).}
\lref\LLi{F.~Livolant, A.M.~Levelut, J.~Doucet and
J.P.~Benoit, Nature {\bf 339}, 724 (1989).}
\lref\RILL{D.~Van Winkle, M.~Davidson and R.L.~Rill,
J. Chem. Phys. {\bf 97}, 5641 (1992); K. Merchant,
Ph.D. Thesis, unpublished.}
\lref\AWM{X. Ao, X. Wen and R.B. Meyer, Physica A {\bf 176},
63 (1991).}
\lref\LLii{A.~Le~Forestier and F.~Livolant, Biophys. J. {\bf 65}, 56 (1994).}

\nfig\fone{A single screw dislocation in a polymer crystal.  The dark
horizontal
line is the screw dislocation.}

\nfig\ftwo{View of a tilt grain
boundary, looking down the $y$-axis.  For clarity we only show the
polymer rows immediately behind and in front of the TGB.  The heavy
lines pointing in the $\hat x$ direction are the screw dislocations.}

\nfig\ffour{Phase diagram of a chiral polymer crystal.  Insets are
representative
tilt (TGB) and moir\'e grain boundaries.  Shaded lines are screw dislocations.
Although
we focus here on TGB and helical moir\'e states, more exotic screw dislocation
phases,
including those with melted dislocation arrays, could also appear.}

\nfig\ffive{Structure function of a tilt grain boundary phase in Fourier space.
Due to the interruption of order by grain boundaries spaced along the $y$-axis
(with period $d'$),
the Bragg spots are broadened along the $q_y$-axis.  This schematic shows a TGB
phase
with
rational rotation angle $\phi=2\pi/7$.}

\nfig\fseven{The moir\'e state.  The thick tubes
running in the $\hat z$ direction are polymers, while the dark lines are
stacked honeycomb arrays of screw dislocations.  The intersection of
these polymers with any constant $z$ cross section away from the hexagonal
defect arrays has the topology of a triangular lattice.}

\nfig\fsix{Structure function of a moir\'e state in Fourier space.  Because
of the periodicity along the $z$-axis, the Bragg spots are broadened along
$q_z$.  This
shows a moir\'e phase with irrational rotation angle
$\phi_3\approx 9.4^{\circ}$ with $10$ crystalline regions.}

\nfig\fnumer{Numerically computed scattering contours in the $q_y$-$q_z$
plane for a moir\'e state with $24$ helical grain boundaries.  We have
scattered from each of $96$
beads evenly spaced along each of the $3721$ polymers.  The structure
function is axially symmetric around the $q_z$ axis.  This numerical
evidence supports the schematic of \fsix .}

\nfig\fnine{A single moir\'e map for triangular lattices with $n=1,2,3,4$
($\phi_n=$ $21.8^{\circ}$, $13.2^{\circ}$, $9.4^{\circ}$, $7.3^{\circ}$
respectively).  The crosses
can be thought of as the ends of polymers in a perfect crystalline region, as
can
the circles.  They join by connecting to the nearest polymer in order to reduce
the
elastic energy.  The shaded lines are the screw dislocations which make up a
honeycomb
network.}

\nfig\fexx{Change in energy from the ground state of an untwisted crystal
as a function of the angle of rotation across a grain boundary, after Balluffi,
Komem
and Schober \BKS .  We expect cusps in the energy around each lock-in angle to
grow
as $\vert\phi-\phi_n\vert$.  In a conventional crystal the cusps are sharper,
behaving
as $\vert\phi-\phi_n\vert\ln\vert\phi-\phi_n\vert$.  Nonetheless, the cusps in
the energy landscape
will be responsible for the lock-in to the special angles $\phi_n$.
}
\nfig\ften{Connectivity diagram for a polymer crystal.  (a) shows the
unperturbed
connectivity at $z=-\infty$.  (b) shows the effect of a single screw
dislocation.  Note
that every vertex has coordination number six on a cross section at
$z=+\infty$,
preserving the nearest neighbor connectivity of the polymers at $z=-\infty$.
(c) shows a section of a
honeycomb
array of screw dislocations near an intersection of three dislocations.  Again
the
coordination number of every vertex is six.}

\nfig\feleven{Allowed moves on a triangular lattice.
Note that the moves
preserve the area between the four vertices.  These moves can be put together
to model the effect of dislocation arrays.}

\nfig\ftwelve{The projected top view of a moir\'e map on a square lattice with
rotation
angle $\tan^{-1}(3/4)$.  The four boxes show the projected polymer
paths after $p=1,2,3,4$ iterations.}

\nfig\fthirteen{A projected top view of $40$ random
polymers paths
resulting from the moir\'e map with $n=1$ iterated 99 times.
There is an exceptional fixed point of all 99 maps at the center.}

\nfig\numform{Numerically calculated average Fourier transform of the monomer
density
of a single polymer in the (a) $q_x$-$q_y$ and (b) $q_y$-$q_z$ planes.
We expect that a dilute concentration of deuterated polymers would produce
neutron scattering profiles similar to these.  The plots were made from
the same data set as in \fnumer .}

\nfig\ffourteen{Radial ($R_r$) and azimuthal ($R_\phi$) radii of gyration near
a supercoincidence site.  The polymers are binned according to $R_0$, the
distance of
their initial point from the center of rotation.  We have rescaled the radii
and $N$
by $R_0$ and collapsed all the data.  The lower curve is
$\log_{10}[R_r^2(N)/R_0]$
and
the upper curve is $\log_{10}[R_\phi^2(N)/R_0] + 2$.}

\nfig\ffifteen{Trajectories of seven nearest neighbors.  The large dots
represent
the starting points of each polymer path.}

\nfig\fsixteen{Radial ($\Delta_r$) and azimuthal ($\Delta_\phi$) root mean
square
separation of nearest neighbors near
a supercoincidence site.  The polymer groups are binned according to $R_0$, the
distance of
their initial point from the center of rotation.  We plot the rescaled
values of $\Delta$ versus $\log_{10}[N/R_0^{0.3}]$.
The lower curve is $\log_{10}[\Delta_r(N)/R_0^{0.24}]$
and
the upper curve is $\log_{10}[\Delta_\phi(N)/R_0^{0.24}] + 2$.}

\Title{IASSNS-HEP-95/10}{\vbox{\centerline{Defects in Chiral Columnar
Phases:}\vskip2pt
\centerline{Tilt Grain Boundaries
and Iterated Moir\'e Maps}}}

\centerline{Randall D. Kamien\footnote{$^\dagger$}{\baselineskip 0.18truein
Address after 1 August 1995: Department of Physics and Astronomy,
University of Pennsylvania, Philadelphia, PA 19104\hfill\break
email: \tt kamien@lubensky.physics.upenn.edu}}
\centerline{\sl School of Natural Sciences, Institute for Advanced Study,
Princeton, NJ
08540}
\centerline{and}
\centerline{David R. Nelson}
\centerline{\sl Lyman Laboratory of Physics,
Harvard University, Cambridge, MA 02138}
\vskip .3truein
\noindent
Biomolecules are often very long with a definite chirality.  DNA, xanthan
and poly-$\gamma$-benzyl-glutamate (PBLG) can all form columnar crystalline
phases.  The chirality, however, competes with the tendency for
crystalline order.
For chiral polymers, there are two sorts of chirality:  the first describes
the
usual cholesteric-like twist of the local director around a pitch
axis, while the second favors the rotation of
the local bond-orientational order and leads to a braiding of the polymers
along an average direction.  In the former case chirality
can be manifested in a tilt grain boundary phase (TGB) analogous to
the Renn-Lubensky phase of smectic-$A$ liquid crystals.  In the latter
case we are led to a new ``moir\'e'' state with twisted bond order.
In the moir\'e state polymers
are simultaneously entangled, crystalline, and aligned, on average, in a common
direction.
In this case the polymer
trajectories in the plane perpendicular to their average direction are
described by
iterated moir\'e maps of remarkable complexity, reminiscent of dynamical
systems.

\Date{20 July 1995}

\newsec{Introduction and Summary}

It is well known that large molecules play a central role in the structure
and function of the cell \AL .  In particular, DNA, large polypeptides such
as poly-$\gamma$-benzyl-glutamate, and polysacharrides such as xanthan
are long polymers with a definite and consistent chirality.
DNA, with a chain length on the order of centimeters, must be packed into
regions with length scales on the order of microns, a scale much smaller than
the average polymer end to end distance in dilute solution.  The packing
of these molecules both {\sl in vivo} and {\sl in vitro} is of great interest.
In the absence of specialized cellular structures, it is plausible
that liquid crystalline phases could facilitate in this packing \ref\LIVO{
F.~Livolant, Physica A, {\bf 176}. 117 (1981).}.
It is known that bacterial plasmids can form both nematic and cholesteric
liquid crystalline mesophases \ref\NATURE{Z.~Reich, E.J.~Wachtel and A.~Minsky,
Science {\bf 264}, 1460 (1994).}.
In addition, dinoflagellate chromosomes \ref\DC{F.~Livolant, Eur. J. Cell Biol.
{\bf 33}, 300 (1984); Y.~Bouligand, M.O.~Soyer and S.~Puiseux-Dao,
Chromosoma {\bf 24}, 251 (1968); R.L.~Rill, F.~Livolant, H.C.~Aldrich and
M.W.~Davidson,
Chromosoma {\bf 98}, 280 (1989).} and bacterial nucleoids \ref\BN{
J.P.~Gourret, Biol. Cell. {\bf 32}, 299 (1978).} exhibit
cholesteric phases, while sperm heads \ref\SH{V.~Luzatti and A.~Nicolaieff, J.
Mol. Biol.
{\bf 1}, 127 (1959); J. Mol. Biol. {\bf 7}, 142 (1963); M.~Feughelman,
R.~Langridge,
W.E.~Seeds, A.R.~Stokes, H.R.~Wilson, M.H.F.~Wilkins, R.K.~Barclay and
L.D.~Hamilton,
Nature {\bf 175}, 834 (1955).} and bacteriophages \ref\BP{J.~Lapault,
J.~Dubochet,
W.~Baschong and E.~Kellenberger, EMBO J. {\bf 6}, 1507 (1987).} exhibit
hexagonal columnar
phases.
Outside the cell, many mesophases arise for long and short
chiral biomolecules: columnar phases \refs{\LLi,\RSDV,\BOUii}, cholesteric
phases
\refs{\RSDV,\RILL,\LLii}, nematic phases
\refs{\Rxi,\Rix,\AWM} and recently
even blue phases \BP\ have been observed.
The possibility of new phases arising
in long chiral molecules is intriguing.

In the hexagonal columnar phase of chiral liquid crystals, the crystalline
close-packing of the molecules
competes with the tendency for the molecules to twist macroscopically around
each other.
As in the twist grain boundary phase of smectic-$A$ liquid crystals \TGB ,
chirality
can enter the crystal through the proliferation of screw dislocations.  In
close
analogy with the physics of type-II superconductors, the screw dislocations
enter
when their energy per unit length is smaller than the energy gain from
introducing chirality.
When the chiral couplings are small, screw dislocations are excluded and a
perfect
equilibrium
crystalline phase persists, in the same way that the Meissner phase expels an
external magnetic field below
the lower critical field $H_{c1}$.

In this paper we elaborate and extend the results presented in \refs{\KN,\KNW}.
As before we neglect structure along the polymer backbones and consider the
allowed
chiral couplings in the hexagonal columnar phase.  We find, in addition to the
usual
cholesteric-like term which favors rotation of the local polymer direction, a
new
coupling which favors the rotation of the crystalline bond order along the
polymer axes.  This
term leads to a novel phase in which the polymers are braided and in which
their trajectories
can be described by iterated moir\'e maps.  Modulated chiral phases have been
discussed in
the context of two-dimensional films \refs{\SL,\PP,\SEL}\ of chiral tilted
molecules.  Our
new phase is a {\sl three-dimensional} chiral modulated structure.  Although
the moir\'e
textures we find have some similarity with the blue phases of
cholesteric liquid crystals, proposed earlier for chiral polymers \LLii , they
differ
in that they {\sl also} incorporate close-packed crystalline order.

We propose two new liquid crystalline phases.  The first is the direct analogue
of the
Renn-Lubensky twist grain boundary phase \TGB\ in which a sequence of polymer
crystal regions
are separated from each other by tilt grain boundaries which effect a finite
rotation of the average polymer direction.  The director ({\sl i.e.} the
polymer
tangent) undergoes a cholesteric-like
rotation around a pitch axis which is perpendicular to the average polymer
direction as well as to the tilt grain boundaries.  We depart here from the
notation
of Renn and Lubensky in calling these walls ``tilt boundaries'' rather than
``twist
boundaries'' as in \TGB .  We use the nomenclature ``helical
grain boundary'' for the honeycomb networks of screw dislocations in the
moir\'e state
discussed now.
The second set of structures are phases characterized
by a sequence of equally spaced parallel helical
grain boundaries which
cause rotations of the local bond-order.  These twist grain boundaries rotate
the bond
order about an axis parallel to the average polymer direction and perpendicular
to the dislocation walls.  This new phase leads to complex polymer
trajectories which are highly entangled and reminiscent of chaotic
dynamical systems.  One or more of these phases should appear in sufficiently
chiral
biopolymers when concentrated in a isotropic solvent.  Additionally, we expect
that discotic liquid crystals, which also form columnar phases, can exhibit
these
phases as well if the molecules are chiral.  It may also be possible to {\sl
induce}
the phases we discuss by imposing appropriately twisted boundary conditions on
weakly chiral or achiral samples.

In section 2 we formulate the theory of chiral polymer crystals, introducing
the
successive degrees of order which separate an isotropic polymer melt from a
directed polymer crystal.
We show how the sequence of phases progresses through spontaneous symmetry
breaking and the non-zero expectations of order parameters.
Additionally, we emphasize the analogy of
crystals to superconductors \HN\ and the concomitant analogy of rotational
invariance with gauge-invariance.

In section 3 we calculate the line energies of various types of dislocations
using continuum elastic theory.  We argue that
DNA as well as discotic
liquid crystals should be in the type-II regime, thus allowing mixed phases
with a proliferation of defects.

In section 4 we analyze the tilt grain boundary (TGB)
phase of the chiral polymers.  Using the continuum elastic theory discussed in
section 3, we
analyze the effect of a wall of screw dislocations on the tilt and twist
fields (nematic director and bond-order).
We estimate the lower critical chiral coupling $\gamma_c$ for the TGB phase in
terms of the Landau parameters.  In addition we show in a continuum
dislocation density
approach
that the TGB phase arises naturally as a low energy configuration of screw
dislocations.  Finally
we discuss the x-ray structure function expected for a macroscopic sample.

In section 5 we analyze the new moir\'e phase.  We
calculate
the lower critical field $\gamma_c'$ for this phase to exist as well as the
effect
on the tilt and twist fields of a single twist wall composed of a honeycomb
network of screw dislocations.  This phase is also shown, in a continuum
``Debye-H\"uckel'' approach,
to arise as a low
energy configuration of screw dislocations.  The braided structure of the
moir\'e
phase contains parallel regions of double-twist \SWM\ as found in the blue
phases of chiral
liquid crystals. Unlike the blue phases, however, the order in most cross
sections perpendicular to the average polymer direction resembles a perfect
triangular lattice.

Finally, in section 6 we propose a microscopic structure for the
moir\'e state.  We model the polymer trajectories as lines woven by a sequence
of moir\'e maps.  We discuss the preferred rotation angles for which moir\'e
patterns
appear as well as the scaling statistics of the polymer trajectories.  Upon
projecting
the polymers onto a plane perpendicular to their average directions,
we study the scaling of the distance
between
neighboring paths, leading to Lyapunov exponents.  In the Appendix we prove
that the special rotation angles that produce moir\'e maps are irrational
fractions
of $2\pi$ and thus the moir\'e state is an incommensurate structure: the
projected polymer
paths never repeat.

\newsec{Rotational Invariance and Free Energies}

We first derive the elastic theory of the polymer crystal, starting from phases
of
higher symmetry.  Columnar crystals are solid-like in two directions and
liquid-like
in one \DGP ;  we assume that the individual monomers in neighboring columns
are out of registry, unlike conventional solids which are crystalline in {\sl
three}
directions.  At high temperatures or in the limit of extreme dilution
the polymers will be isotropically entangled.
We proceed by breaking successive rotational and translational symmetries until
a hexagonal columnar phase is reached.  All these phases need not exist for a
given
material; one or more of the intermediate phases could be skipped via a direct
first
order transition.

\subsec{The Polymer Nematic Phase}
When
the
temperature $T$ is decreased or the density increased, an isotropic polymer
melt can transform into
a nematic phase
\ref\BKZ{L.~Balents, R.D.~Kamien,
P.~Le~Doussal and E.~Zaslow, J. Phys. I (Paris) {\bf 2}, 263 (1992).}.
Orientational
fluctuations are then described by the usual Frank free energy density:
\eqn\efrank{\eqalign{\free_{{\bf n}} &= \half \left\{K_1(\nabb\dot{\bf n})^2
+K_2[{\bf n}\dot(\nabb\cross{\bf n})]^2 + K_3[{\bf n}\cross(\nabb\cross{\bf
n})]^2\right\}
\cr&\approx\half\left[K_1(\grad\dot\dnb)^2 + K_2(\grad\cross\dnb)^2
+K_3(\dz\dnb)^2\right],\cr} }
where the $\{K_i\}$ are the splay, twist and bend elastic constants.  In the
nematic phase we can take ${\bf n}\approx {\bf\hat z} + \dnb$ where $\dnb$ is a
vector
in the plane perpendicular to the average director.  Here, and in the
following, vectors which lie in the plane perpendicular to the average
direction of the director will be denoted as $\vec A$, while full
three dimensional vectors will be represented by bold face $\bf A$.  We
will henceforth take the average nematic direction to be $\bf\hat z$.
Because the polymers
are long, $K_1$ is much larger than it would be in a comparable short-chain
liquid
crystal \refs{\MEYER,\KLN}, diverging when the polymers become infinitely long.
In addition, non-linear effects are known to modify the behavior of $K_2$ and
$K_3$ and the bulk compression modulus, leading to their anomalous dependence
on the polymer lengths \refs{\TONER,\KT} .
If the
system were chiral, terms which violate parity transformation would be allowed.
 In a
nematic phase
the director has a discrete ${\bf n}\rightarrow-{\bf n}$ symmetry which must
be preserved in the free energy.
Since under parity
$\nabb\rightarrow -\nabb$, the lowest order chiral term is
${\bf n}\dot(\nabb\cross{\bf n})$.  Note that since every term must have
even powers of $\bf n$ we can chose to take ${\bf n}\rightarrow -{\bf n}$
(vector)
or ${\bf n}\rightarrow{\bf n}$ (pseudovector) under parity.  For definiteness
we take $\bf n$ to be a vector, though this has no effect on the analysis or
results.  We write the chiral contribution to the nematic free energy as
\eqn\echol{\free^*_{{\bf n}} = -\gamma{\bf n}\dot(\nabb\cross{\bf n})\approx
-\gamma\grad\cross\dnb,}
where the two dimensional pseudoscalar cross product is $\vec a\cross\vec b =
\epsilon_{ij}a_ib_j$.
One ground state configuration of $\free_{{\bf n}}+\free^*_{{\bf n}}$ is a
cholesteric state with pitch $q_0=\gamma/K_2$, although more exotic blue phases
are
also
possible \SWM .  Nonlinearities associated with
rotational invariance are known to modify the relation between $\gamma$ and
$q_0$ \KT .

\subsec{Nematic-hexatic (N+6) Phase with Chirality}
At still lower temperatures or higher concentrations a new phase may occur
which includes hexatic order
in the plane perpendicular to the nematic director ${\bf n}$ \TON .
Although the present experimental evidence for the hexatic order in polymer
nematics is sketchy, such phases seem highly likely in columnar systems, in
analogy with
the hexatic order possible for vortex lines in high temperature
superconductors \MN .
There is a complex order
parameter, $\psi_6$ which characterizes the local bond order.
Its behavior is determined by a Landau theory
\eqn\lansix{\free_6= {1\over 2}h_1\vert{\bf n}\dot\nabb\psi_6\vert^2
+{1\over 2}h_2\vert\nabb\psi_6\vert^2 + {r\over 2}\vert\psi_6\vert^2
+{u\over 2}\vert\psi_6\vert^4}
with $r=a(T-T_6)$.  In the
ordered phase, for $T<T_6$, $\langle\psi_6\rangle
\neq 0$ and we may write $\psi_6=\vert\psi_6\vert e^{6i\ts}$.
Sufficiently
below the transition we may neglect the fluctuations in the magnitude of the
order parameter and employ a Landau theory for $\theta_6$.  Upon
suppressing an additive constant, $\free_6$ becomes
\eqn\ehex{\free_{6} = \half K_A^{||}({\bf n}\dot\nabb\theta_6)^2 + \half
K_A^{\perp}
\left[(\nabb\ts)^2 -({\bf n}\dot\nabb\theta_6)^2\right]\approx \half
K_A^{||}(\dz\ts)^2
+\half K_A^{\perp}(\grad\ts)^2}
where $K_A^{||}=36\vert\psi_6\vert^2(h_1+h_2)$ and
$K_A^{\perp}=36\vert\psi_6\vert^2h_2$
are hexatic stiffnesses parallel and perpendicular
to the nematic director and $\vert\psi_6\vert^2 = a(T_6-T)/4u$.
The symmetry ${{\bf n}}\rightarrow -{{\bf n}}$ discussed above for pure
nematics
must also hold in this phase. However, since
$\ts$ is measured around ${{\bf n}}$, upon taking ${{\bf n}}\rightarrow -{{\bf
n}}$
we must
also have
$\ts\rightarrow-\ts$.  If, for instance, we had originally measured a change in
$\ts$
with respect to the right-hand rule using ${{\bf n}}$, we would now, see the
change
go in the opposite direction, with respect to $-{{\bf n}}$.  Thus we may have
any
term
quadratic in powers of $\ts$ or $\bf n$.  Under parity, $\theta_6$ will
change sign
if we take ${\bf n}$ to transform as a vector, for the same reason it changes
sign
under nematic inversion. Note that $\ts\rightarrow-\ts$ is equivalent to
$\psi_6\leftrightarrow\psi_6^*$.  We are now able to write a new chiral term
\KNT\
for the ``N+6'' phase.  Namely
\eqn\ehexc{\free^*_6 = -i{\Gamma'\over 12}{\bf n}\dot\left(\psi_6^*\nabb\psi_6
-
\psi_6\nabb\psi_6^*\right) = -\gamma' {\bf n}\dot\nabb\theta_6\approx
-\gamma'\dz\ts}
where $\gamma'=\Gamma'\vert\psi_6\vert^2$. In a nematic ground state with ${\bf
n}=\bf\hat z$ everywhere ({\sl i.e.} $\gamma\approx 0$),
this term will favor a ground state with $\theta_6$ twisting with a pitch
$\tilde q_0=\gamma'/K_A^{||}$ along the $z$-axis.  A similar coupling has
been considered before in hexatic smectic-$B$ phases \DGP .
If both $\gamma$ and $\gamma'$ are nonzero and $q_0$ and $\tilde q_0$ are
incommensurate,
one might expect that a chiral N+6 polymer melt would resemble
an incommensurate smectic \refs{\TONP,\KNT}.  These considerations also
apply
to short chain nematogens with the same symmetries:
if a chiral nematic (or discotic) had the analogue of an N+6 phase {\sl two}
distinct chiral couplings would again be allowed.
Table 1
gives a
list of important couplings with their symmetry properties.

There are, in addition to the chiral coupling between $\bf n$ and $\ts$,
non-chiral
couplings between $\bf n$ and $\ts$ \refs{\TON,\GIA}.  Based on rotational,
nematic and parity
invariance, the lowest order terms are
\eqn\efnvi{\free_{{\bf n}6} = \bar C({\bf n}\dot\nabb\ts)[{\bf
n}\dot(\nabb\cross{\bf n})] + \bar C'\nabb\ts\dot(\nabb\cross{\bf n})\approx
C\dz\ts\grad\cross\dnb + C'\grad\ts\cross\dz\dnb}
where $C=\bar C +\bar C'$ and $C'=\bar C'$.
While the final two terms differ only by a total derivative, they are, in
principle, different in the presence of topological defects in $\bf n$ and
$\ts$.

\subsec{Hexagonal Columnar Phases with Chirality}
At still lower temperatures $T<T_{\rm xtal}$ (or higher concentrations)
the polymers can crystallize in
the
$xy$-plane.  To insure rotational invariance, and, in analogy with
treatments of crystals
made up of point particles \NT , we build up the crystal out of a superposition
of plane density waves with wavevectors ${{\bf G}}_\alpha$.
In the case of a triangular lattice we could take,
for example, ${{\bf G}}_\alpha=\vert {\bf G}\vert[\cos(\pi\alpha/3),
\sin(\pi\alpha/3),0]$ where $\alpha=1,\ldots,6$, $\vert{\bf G}\vert =
4\pi/(\sqrt{3}a_0)$
and $a_0$ is the lattice constant of the crystal.
The $\{{{{\bf G}}}_\alpha\}$ are the six smallest
reciprocal lattice vectors of a triangular lattice and ${\bf
G}_\alpha\dot\bf\hat z =0$.
The areal polymer density in a plane perpendicular to the average
direction
may be expanded as
\eqn\waves{\rho \approx \rho_0 + \sum_{\alpha=1}^6 \rho_\alpha(
{{\bf r}}) \exp\{-i{\bf G}_\alpha\dot{\bf r}\}}
where higher order reciprocal lattice vectors could also be included.

Each plane wave is modulated by a spatially varying magnitude and
phase $\rho_\alpha({{\bf r}}) = \vert\rho_\alpha({{\bf r}})\vert{\exp\{i{{\bf
G}_\alpha}\dot
{\vec u}({{\bf r}})\}}$, where $\vec u$ is a two-dimensional displacement
field.
Under a rotation by $\bold{\theta}$, the plane waves in \waves\ change.  In
particular
${\bf r}\rightarrow{\bf r}+\bold{\theta}\times{\bf r}$, and so ${\bf
G}\dot{\bf r}\rightarrow
{\bf G}\dot{\bf r} + {\bf r}\dot\left({{\bf G}}\cross\bold{\theta}\right)$.
Thus
rotations lead to a position dependent change in $\rho_\alpha({\bf r})$, namely
$\rho_\alpha\rightarrow\rho_\alpha\exp\{i{{\bf G}}\dot(\bold{\theta}\cross{\bf
r})
\}$.
Likewise, under a global rotation about the $x$-axis or
$y$-axis by an angle $\theta_x$ or $\theta_y$ respectively, ${\bf n}\rightarrow
{\bf n} +\bold{\theta}\times{\bf n}\approx {\bf\hat z} +\theta_y{\bf\hat x}
-\theta_x{\bf\hat y}$, {\sl
i.e.} $\dnb \rightarrow \dnb + \theta_y{\bf\hat x} -\theta_x{\bf\hat y}$.
Similarly,
under a global rotation about the $z$-axis by $\theta_z$,
$\ts\rightarrow\ts+\theta_z$.
To insure rotational invariance \refs{\HN,\NT}, derivatives of
$\rho_\alpha({\bf r})$
must be accompanied by the fields $\ts$ and $\dnb$ .
\eqn\efree{\eqalign{
&\free_{\rm xtal}=\cr&\;\sum_\alpha\bigg\{
{A\over 2}\left\vert{{\bf G}}_\alpha\dot\left[
\vec\nabla_{\!\perpp}\rho_\alpha - i\ts\left({{\bf G}}_\alpha\cross{\bf\hat
z}\right)\rho_\alpha\right]\right
\vert^2
+{B\over 2}\left\vert{{\bf G}}_\alpha\cross\left[
\vec\nabla_{\!\perpp}\rho_\alpha - i\ts\left({{\bf G}}_\alpha\cross{\bf\hat
z}\right)\rho_\alpha\right]\right
\vert^2
\cr
&\;+{C\over 2}\left\vert
\partial_z\rho_\alpha - i\left({\bf
G}_\alpha\dot\dnb\right)\rho_\alpha\right\vert^2
+ {b\over 2}\vert\rho_\alpha\vert^2\bigg\}+c\sum_{\alpha\beta\gamma}^{{\bf
G}_\alpha
+{{\bf G}}_\beta +{{\bf G}}_\gamma = \bf 0}
\rho_{\alpha}\rho_{\beta}\rho_{\gamma} +
\bo{\rho_\alpha^4}\cr}}
Crystalline order arises in this Landau expansion for sufficiently small
$b\sim
(T-T_{\rm xtal})$ so
that
$\langle\,\rho_\alpha\,\rangle\ne 0$ .  Due
to the third order term, this transition will, in general, be first order,
and $T_{\rm xtal}$ represents a limit of metastability.
The free energy is the sum of all the terms discussed earlier, namely
\eqn\ethefr{F=\int d^3\!x\,\left\{\free_{\bf n}+\free_6 + \free_{{\bf n}6} +
\free^*_{\bf n}
+\free^*_6 + \free_{\rm xtal}\right\}}

The free energy \ethefr\ is analogous to the Ginzburg-Landau theory of a
superconductor.
Here, rotational invariance dictates the couplings between Goldstone modes,
$\dnb$
and $\ts$, and derivatives of $\rho_\alpha$ in the same way that gauge
invariance
dictates the coupling between derivatives of the BCS order parameter $\psi$ and
the electromagnetic field $\bf A$.  The contribution $(\free_{\bf n} +\free_6
+\free_{{\bf
n}6})$ mimics the field energy $(\curl{\bf A})^2$.  The two chiral couplings
are analogous to the coupling between the magnetic field ${\bf
B}\equiv\curl{\bf A}$ and the
external field $\bf H$ \DGS .  In our case there are {\sl two} distinct
``magnetic
fields'' $\bf H$ represented by the chiral couplings $\gamma$ and $\gamma'$.

Upon setting $\rho_\alpha = \vert\rho\vert\exp\{i{{\bf G}}_\alpha\dot\vec u\}$
and
integrating out $\dnb$ and $\ts$ we find that these fields
become locked to various derivatives of the
displacement fields:
\eqn\estarthing{\eqalign{\delta n_i &\approx \partial_z u_i\cr
\ts&\approx\half\epsilon_{ij}\partial_iu_j\cr}}
The resulting free energy to lowest
order in $\vec u$ and its derivatives is now
\eqn\efreeatlast{F=\int d^3\!x\,\left\{\mu u_{ij}^2 + {\lambda\over 2}u_{ii}^2
+
K_3(\partial_z^2u_i)^2 -\gamma\vec\nabla_{\!\perpp}\cross\dnb
-\gamma'\partial_z\ts\right\} }
where $u_{ij} =\half(\partial_iu_j+\partial_ju_i)$.  Here and throughout
Roman indices indicate directions only in the $xy$-plane while Greek indices
indicate all three coordinate directions.  In terms of the
parameters in \efree , $\mu = {3\over 4}\vert{\vec
G}\vert^4\vert\rho\vert^2(A+B)$ and
$\lambda={3\over 4}\vert{{\bf G}}\vert^4\vert\rho\vert^2(A-B)$.  In contrast
with polymer
nematics, in the crystal the elastic constants do not diverge when to the
nonlinearities
associated with rotational invariance \KNT\ are taken into account.  The
simplified free
energy \efreeatlast\ is similar to the Ginzburg-Landau theory in the London
limit.

The response of superconductors to an external magnetic field is determined by
$\kappa$,
the ratio of the penetration length to the coherence length.  As discussed in
the
next section, the parameter analogous to $\kappa$ in hexagonal columnar
crystals is
typically much greater than $1$, showing that these materials behave as Type II
rather
than Type I superconductors in their response to chirality.  In other words the
equilibrium ground states may contain defects.

\newsec{Isolated Dislocations and Their Energies}

When dislocations are introduced into the free energy \efreeatlast\
$\vec u(\bf r)$ is no longer
single valued.  To account for this, we introduce a new variable
$w_{\gamma i}$ which is equal to $\partial_\gamma u_i$ away from the
defects \KOS .
The free energy becomes
\eqn\efreeii{
F=\int d^3\!x\,{\mu}\left({w_{ij}+w_{ji}\over 2}\right) + {\lambda\over
2}(w_{ii})^2+{K_3\over 2}(\dz w_{zi})^2
- \gamma\epsilon_{ij}\di w_{zj} - \gamma'\dz({1\over
2}\epsilon_{ij}w_{ij}),}
where $\ts={1\over 2}\epsilon_{ij}w_{ij}$ and $\dn_i=w_{zi}$.

Dislocations are restricted so that the Burger's
vector, $\vec b$ must lie
in the $xy$ plane and ${\bf t}\dot({\bf n}\cross {\vec b})=0$, where $\bf t$ is
the unit tangent point along the dislocation line.  The latter constraint
eliminates dislocations
which add a row of polymer ends \MN , which we neglect.
We introduce the density tensor $\alpha_{\gamma i}({\bf r}) = \int d{\bf t}
d{\vec b}\,
t_\gamma b_i \rho({\bf t},\vec b, {\bf r})$, where $\rho({\bf t},\vec b,{\bf
r})$ is the volume density
of dislocations at the point ${\bf r}$ with Burger's vector $\vec b$ pointing
in
the
$\bf t$ direction. Since the dislocations do not end, $\nabb\cdot {\bf t}=0$,
and
$\partial_\gamma\alpha_{\gamma i}\equiv 0$.
The constraint $0={\bf t}\dot({\bf n}\cross\vec b) = {\bf n}\dot
(\vec b\cross{\bf t})$ becomes, assuming ${\bf n}||\bf\hat z$, $b_xt_y=b_yt_x$
or
in the case of many defects $\alpha_{xy}=\alpha_{yx}$.  In terms of $w_{\gamma
i}$
this constraint reads $\partial_i w_{zi} = \partial_z w_{jj}$, {\sl i.e.}
$\grad\dot\dnb = -\dz w_{jj}\equiv
-\dz(\delta\rho/\rho)$ ($\delta\rho$ is the fluctuation in the areal density of
polymers $\rho$ in a constant-$z$ cross section)
which creates the sort of long-range interactions that
are central to theories of directed polymers \refs{\TM,\SB,\KLN}.

Following reference \KOS , we can relate $w_{\gamma i}$ to the density of
dislocations inside a small area $\Gamma$ as follows:
\eqn\eburgi{\eqalign{
\oint_{\partial\Gamma} ds\, {du_i\over ds} &= -\sum_n b^n_i\cr
\oint_{\partial\Gamma} dx_\gamma\, {du_i\over dx_\gamma}
&= -\sum_n\int_\Gamma {\bf t}^n\cdot d{\bf S}\,b^n_i\delta^2({\bf
t^n\cross r})\cr
\int_\Gamma dS_\mu\,\epsilon_{\mu\nu\gamma}\partial_\nu w_{\gamma i} &=
-\int_\Gamma
dS_\mu \alpha_{\mu i}\cr}}
where $\partial\Gamma$ is the boundary of $\Gamma$ and $d{\bf S}$ is a directed
surface element.  It follows that
\eqn\eburgii{
\epsilon_{\mu\nu\gamma}\partial_\nu w_{\gamma i}({\bf r}) = - \alpha_{\mu
i}({\bf r}).}
In Fourier space \eburgii\ can be solved for $w_{\gamma i}$ and we have
\eqn\wdoubleu{w_{\gamma i}({\bf q}) = -i{\epsilon_{\gamma\mu\nu}
q_\mu\alpha_{\nu
i}({\bf q})\over q^2}
+ iq_\gamma\psi_i({\bf q}),}
where it is convenient to decompose the arbitrary function $\psi({\bf q})$ into
longitudinal
and transverse parts,
\eqn\epsi{\psi_i=i{q_i\sigma\over q_\perp^2} - i{\epsilon_{ij}q_j\pi\over
q_\perp^2}}
so that $\grad\!\cdot\psi({\bf r})=-\sigma$ and $\grad\!\times\psi({\bf r}) =
-\pi$.
The field $\vec\psi$ describes the equilibrium displacement
of the polymers in the presence of the dislocation density $\alpha_{\gamma
i}({\bf r})$.
The extremal
equations which result from minimizing \efreeatlast\ with respect to variations
in $\vec\psi$
imply that
\eqn\divpsi{[(2\mu +\lambda)q_\perp^2 + K_3q_z^4]\vec q_\perp\dot\vec\psi =
-{2\mu+\lambda-K_3q_z^2\over q^2} \epsilon_{lm}
q_zq_lq_j\alpha_{mj}}
and
\eqn\curlpsi{[{\mu}q_\perp^2 + K_3q_z^4]\vec q_\perp\cross\vec\psi=
-{{\mu}-K_3q_z^2\over q^2}q_zq_sq_i\epsilon_{sj}\epsilon_{im}\alpha_{mj}
+{\mu+\lambda\over q^2} q_sq_iq_\rho\epsilon_{sj}\epsilon_{j\rho\nu}\alpha_{\nu
i}.}
The three terms on the rights hand sides
are sources representing the three different types of dislocation.
An edge
dislocation lying in the $xy$-plane provides a source for \divpsi .
The
first source contribution
in \curlpsi\ is the only non-vanishing term for a straight screw
dislocation, while the second
is non-vanishing for edge dislocations which lie parallel to the $z$-axis.

Following \IND , we can find the effect of dislocations
on the the nematic twist
$\grad\cross\dnb$ and bond twist $\dz\ts$.  Let
$\omega_\rho=\epsilon_{\rho\alpha
j}
w_{\alpha j}$.  Then, \wdoubleu\ leads to $\nabb\dot\bold{\omega} =
-\Tr[\alpha]$.
The definitions $\ts={1\over 2}\epsilon_{ij}w_{ij}$ and $\delta
n_i = w_{zi}$ also imply $\nabb\dot\bold{\omega} =
\partial_y\dn_x-\partial_x\dn_y  +
2\dz\ts$.
Thus
\eqn\ior{2\dz\ts-\grad\cross\dnb = -\Tr[\alpha].}
Since the diagonal components of $\alpha_{ij}$ represent the density of screw
dislocations
we have related the quantities we are interested in to the macroscopic defect
density.
Upon referring to \estarthing , we see that \ior\
quantifies the non-commutivity of
derivatives
of $\vec u$.  By choosing different dislocation complexions we can make one or
both
of $\dz\ts$ and $\grad\cross\dnb$ non-zero and thus take advantage of the
chiral couplings
in \efreeatlast .

We can now solve for the strain field around a screw dislocation.
A screw dislocation in a polymer crystal is shown in \fone, where we have
chosen
boundary conditions such that $\dnb\rightarrow\vec 0$ at $z=\pm\infty$.  For a
screw dislocation
along
the $x$-axis,
$\alpha_{\gamma i} = \delta_{\gamma x}\delta_{ix}\delta(y)\delta(z)b$
or, in Fourier space, $\alpha_{\gamma i} = 2\pi b\delta_{\gamma
x}\delta_{ix}\delta(q_x)$.
Immediately from \wdoubleu\ we see that $w_{xx}=w_{xy}=0$. Since
$\vec q_\perp\dot\vec \psi=0$
we also have $w_{yy}=w_{zy}=0$.
It follows that
\eqn\ewyx{w_{yx}=-{2\pi i\delta(q_x)bq_z\over q^2}
\Big[1 - {q_y^2({\mu}-K_3q_z^2)\over{\mu}q_\perp^2+K_3q_z^4}\Big]}
and
\eqn\ewzx{w_{zx}={2\pi i\delta(q_x)bq_y\over q^2}
\Big[1+{q_z^2({\mu}-K_3q_z^2)\over{\mu}q_\perp^2+K_3q_z^4}\Big].}

{}From these results we can readily determine the bond twist and nematic twist
fields generated
by a single screw dislocation with the boundary conditions as above.  The bond
twist is
\eqn\dzts{\eqalign{
\dz\ts &= -{1\over 2}\dz w_{yx}=
-{K_3b\over 2}\int {dq_ydq_z\over (2\pi)^2}\,{q_z^4\over{\mu\over
2}q_y^2+K_3q_z^4}e^{i(q_yy+q_zz)}\cr
&= -{b\over 16\sqrt{\pi}}\lambda^{-1/2}\vert y\vert^{-5/2}e^{-{z^2\over
4\lambda\vert y\vert}}\left\{\vert y\vert-{z^2\over 2\lambda}\right\}\cr}}
where
\eqn\eladda{\lambda=\sqrt{K_3\over \mu}.}
The nematic twist is
\eqn\curln{\eqalign{\grad\cross\dnb &= \dx w_{zy} - \dy w_{zx}={\mu b\over
2}\int {dq_ydq_z\over (2\pi)^2}\
{q_y^2\over{\mu\over 2}q_y^2+K_3q_z^4}e^{i(q_yy+q_zz)}\cr
&=b\delta(y)\delta(z)+2\dz\ts\cr}}
Note that \dzts\ and \curln\ are related as in \ior .

Boundary conditions
play a crucial role in determining the integrated value of the twists.
The key issue is whether the boundaries at large $\vert z\vert$ go
to infinity before or after the boundaries at large $\vert y\vert$.  It is
enlightening
to consider, for instance, a vortex disclination in a two-dimensional XY model.
 While
a symmetric configuration with the spins pointing out radially
from the core may have the lowest elastic energy, boundary conditions can
change
this configuration.  In an $L\times L$ region of
the $xy$-plane we could impose boundary conditions such that
the spins are normal to the surfaces at $x=\pm L$.  All the winding
of the order parameter at the boundaries will occur along the lines at $y=\pm
L$.  In the case of a dislocation
in the columnar crystal, either $\ts$ or $\dnb$ ``wind'', depending on the
boundary
conditions.   The boundary
condition $\dnb\rightarrow\vec 0$ as $z\rightarrow\pm\infty$ implies that there
is no net twist of the nematic director.  In the above example and in \fone\ we
have
enforced this constraint on
surfaces at $z=\pm L$ for all values of $y$.  This is equivalent to
taking the large $y$ cutoff to infinity first, {\sl i.e.} $q_y\rightarrow 0$.
Hence
in this case, the integral over a constant $x$ plane of $\dz\ts$ is
\eqn\eallspdzts{\int dydz\,\dz\ts({\bf r}) = \lim_{q_z\rightarrow 0}
\lim_{q_y\rightarrow
0} \dz\ts({\bf q}) = -{b\over 2},}
while the similar integral of $\grad\cross\dnb$ is
\eqn\eallspcurln{\int dydz\,\grad\cross\dnb({\bf r}) = \lim_{q_z\rightarrow 0}
\lim_{q_y\rightarrow
0} \grad\cross\dnb({\bf q}) = 0.}
The boundary conditions and the order of limits
will be of importance when adding up the effects of a collection of defects, as
we shall see in
the following sections.

Upon inserting the expressions for $w_{\gamma i}$ into \efreeii , we find the
screw
dislocation energy per unit length
\eqn\edefect{\eqalign{f_{\rm screw} &= {1\over L}\int {d^3\!q\over (2\pi
)^3}\,\left[{\mu\over 2}\vert w_{yx}\vert^2
+ {K_3\over 2}\vert q_z
w_{zx}\vert^2\right] = {\mu K_3b^2\over 2(2\pi)^2}\int
dq_ydq_z\,{q_z^2\over{\mu}q_y^2+K_3q_z^4}\cr
&={\mu^{3/4}K_3^{1/4}b^2 \over 2\pi^2\sqrt{\xi_\perp}}
\left\{\eqalign{&\delta\arctan\left({1\over\delta^2}\right)+{1\over
2\sqrt{2}}\ln\left(
{\delta^2-\sqrt{2}\delta+1\over\delta^2+\sqrt{2}\delta+1}\right) \cr&\qquad+
{1\over\sqrt{2}}
\arctan\left({\sqrt{2}\delta\over 1-\delta^2}\right)\cr}\right\}\cr},}
where
\eqn\egamma{\delta=\left({K_3\over\mu}{\xi_\perp^2\over\xi_z^4}\right)^{1/4},}
and $\xi_\perp$ and $\xi_z$ are the short-distance $y$ and $z$ cutoffs,
respectively.
In the two extreme limits, the elastic free energy per unit length becomes
\eqn\elimits{
f_{\rm screw} = \left\{\eqalign{{\mu^{3/4}K_3^{1/4}b^2\over 2\sqrt{2}\pi
\sqrt{\xi_\perp}}&\qquad\delta\rightarrow\infty\cr
{\sqrt{\mu K_3}b^2\over 4\pi\xi_z}&\qquad\delta\rightarrow 0\cr}\right.}
so the deciding factor is the dimensionless parameter $\delta$.
In analogy with superconductors, $\delta^2$
plays the role of $\kappa$, the ratio of the penetration depth to the
coherence length.  Indeed,
suppose for simplicity that the two cutoffs $\xi_\perp$ and $\xi_z$
are
comparable ($\xi_{\perp}\sim\xi_z\equiv\xi)$ so that $\delta^2 \approx
\sqrt{K_3/\mu}/\xi$.
{}From
inspection of \efreeatlast , we see that $\lambda=\sqrt{K_3/\mu}$ is the length
scale over which bend deformations heal. so
$\sqrt{K_3/\mu}$ plays the role of the London penetration depth in a
superconducting
analogy \DGP .  Thus $\delta^2$ is the ratio of the
healing length of director fluctuations over the healing length
of density fluctuations ($\xi\sim a_0$, the lattice constant) which
is precisely the form of $\kappa$, in superconductors.  When $\delta\ll 1$ the
polymer
crystal is
type-I and chirality will be excluded until the crystal breaks down
completely forming a chiral liquid.  However, when $\delta\gg 1$ type-II
behavior
occurs and chirality can creep into the crystal through the proliferation of
defects.  We must determine the magnitude of $\delta$ to decide which sort
of behavior we expect.

We first consider chiral polymers, such as DNA.  We take $\xi_\perp$ to be
the
average spacing $a_0$
between the polymer strands and $\xi_z$ to be the on the
order of the polymer diameter, or alternatively the base pair
spacing in DNA.
Then $K_3=\kbT L_p/\xi_\perp^2$
and $\mu\approx \epsilon_0/\xi_\perp^2$, where $L_p= \kappa_b/\kbT$  is the
persistence
length ($\kappa_b$ is the polymer bending stiffness)
and $\epsilon_0$ is the polymer interaction energy per unit
length.
With the experimental parameters in \PARS , we have
$\xi_z\approx 5\angstrom$, $\xi_\perp=30\angstrom$, $L_p=550\angstrom$
and $\epsilon_0 \approx 1.5\kbT\angstrom^{-1}$ leading to
$\delta\approx 5$.  Thus we expect type-II behavior.

For dense chiral discotics, we take $\xi_\perp$ and $\xi_z$ to be the
respective sizes of the disk-shaped molecules.  In this case we expect
that $K_3\sim U_0/\xi_z$ and $\mu\sim U_0/(\xi_\perp^2\xi_z)$ where $U_0$ is a
characteristic interaction energy.
If the interactions
were entirely steric we would expect that $U_0\sim \kbt$.  Using \egamma\ we
then find that
$\delta = \xi_\perp/\xi_z$.  In discotics, as the name implies, $\xi_\perp\gg
\xi_z$ and so we expect that they will also be type-II columnar crystals.  Note
that in the limit of rod like-molecules (as in nematics) $\xi_\perp\ll\xi_z$
and we expect type-I behavior in the hypothetical situation of
nematic molecules forming a hexagonal columnar phase.

Straightforward estimates, along the lines taken in \ref\VAREN{D.R.~Nelson,
in {\sl Observation, Prediction and Simulation of Phase Transitions in
Complex Fluids}, edited by M.~Baus, L.F.~Rull and J.-P.~Ryckaert (Kluwer,
Dordrecht, 1995).} ,
show that the screw dislocation core energy per unit length, when
the magnitude of the Burgers vector $b=a_0$ and $\delta\gg 1$, is
\eqn\ecorestu{E_c = cK_3\mu^{3/4}a_0^{3/2}}
where $c$ is a numerical constant of order unity.  Thus $E_c$ is the same
order of magnitude as the $\delta\rightarrow\infty$ limits of the elastic
energy displayed in
\elimits .

Considering now an edge dislocation with a Burgers vector of magnitude $b$ and
parallel
to the $z$ axis
$\alpha_{\gamma i} = \delta_{\gamma z}\delta_{ix}b\delta(x)\delta(y)$, and
the free energy per unit length is
\eqn\eedge{f_{\rm z-edge} = {\mu b^2\over 2(2\pi)^2}\int dq_xdq_y\,
{q_y^6+2q_x^6+5q_x^4q_y^2+4q_x^4q_y^2\over q^8}={3\mu b^2\over
16\pi}\ln\left({R\over\xi}\right)}
which, unlike the screw dislocation energy,
will diverge as the logarithm of the system size.  The core
energy is $E_c=c'\mu b^2$ in this case.  The logarithmic divergence arises
because a vertical edge defect acts on each $xy$-layer as a normal
edge dislocation in a two-dimensional crystal.

Unlike infinitely long polymers, discotics can also
have edge dislocations lying in the $xy$-plane.
For an edge dislocation running along the $y$ axis, we have
$\alpha_{\gamma i}=
b\delta_{\gamma y}\delta_{ix}\delta(x)\delta(z)$ and find an elastic free
energy per
unit length
\eqn\ebadedge{f_{\rm xy-edge} ={(2\mu+\lambda) K_3b^2
\over 2(2\pi)^2}\int dq_xdq_z\,{q_z^2\over(2\mu+\lambda) q_x^2+K_3q_z^4}}
which is finite, and can be reduced to the complicated expression \edefect\ by
replacing $\mu$ with $(2\mu+\lambda)$ \DGP .

\newsec{The Tilt Grain Boundary Phase}
\subsec{Strains, Energies and Displacements}
The twist grain boundary state was first proposed as the
analogue of the Abrikosov flux lattice in chiral smectic-$A$ liquid crystals
\TGB\ and
later discovered to exist in nature  \ref\TBBe{J. Goodby, M.A. Waugh, S.M.
Stein, R. Pindak, and J.S. Patel, Nature {\bf 337},
449 (1988); J. Am. Chem. Soc. {\bf 111}, 8119 (1989); G. Strajer, R. Pindak,
M.A. Waugh, J.W.
Goodby, and J.S. Patel, Phys. Rev. Lett {\bf 64}, 13 (1990); K.J. Ihn, J.A.N.
Zasadzinski, R.
Pindak, A.J. Slanet, and J. Goodby, Science {\bf 258}, 275 (1992).}.  In
analogy with this
state, and, as suggested by Kl\'eman \KLEMi , the tilt grain boundary phase of
polymers
is a crystalline version of a polymer cholesteric, where the cholesteric twist
arises from a sequence of low-angle tilt grain boundaries.

We first consider a single grain boundary.  The wall is made up of screw
dislocations,
pointing
along the $\hat x$ direction, stacked up in the $xz$-plane.  An example with
three
screw dislocations is shown in
\ftwo .
To determine the twists and strain, we superpose the strains from
each individual screw dislocation.  With screw dislocations at $z=0,\pm d,\pm
2d\ldots$,
we have
\eqn\tgbcurl{
[\grad\cross\dnb]_{\rm TGB} = \sum_{n=-\infty}^{\infty}
[\grad\cross\dnb(x,y,z-nd)]_1}
where the notation $[\ldots]_1$ refers to the result \curln\ for a single screw
dislocation.
With the help of the Poisson summation formula \tgbcurl\ can be rewritten as
mixed Fourier transform
\eqn\tgbcurlii{\eqalign{
[\grad\cross\dnb]_{\rm TGB}&= {1\over d}\sum_{m=-\infty}^\infty e^{2\pi imz/d}
[\grad\cross\dnb(x,y,q_z={2\pi
m\over d})]_1 \cr
&={b\over d}\delta(y)\sum_m e^{2\pi imz/d}
-{(2\pi)^2b\lambda\over 2d^3}\sum_m e^{2\pi imz/d}
m^2e^{-\lambda(2\pi)^2m^2\vert
y\vert/d^2}\cr},}
where \curln\ has been used in the second line.
Since $d$ is small compared to $y$ and $z$ in the far field limit,
we approximate the second sum by the first term, the remaining terms
being exponentially suppressed:
\eqn\tgbcurliii{\eqalign{[\grad\cross\dnb]_{\rm TGB}&\approx
b\delta(y)\sum_{n=-\infty}^{\infty}\delta(z-nd) - {(2\pi)^2b\lambda\over
d^3}\cos\left({2\pi z\over
d}\right)e^{-(2\pi)^2\lambda\vert y\vert/d^2}\cr
&\approx{b\over d}\delta(y)- {(2\pi)^2b\lambda\over d^3}\cos\left({2\pi z\over
d}\right)e^{-(2\pi)^2\lambda\vert y\vert/d^2}\cr}.}
Similar manipulations lead to
\eqn\tgbdzts{[\dz\ts]_{\rm TGB} \approx - {(2\pi)^2b\lambda\over
2d^3}\cos\left({2\pi z\over
d}\right)e^{-(2\pi)^2\lambda\vert y\vert/d^2},}
and again \ior\ is satisfied, since now the density of dislocations is
precisely
$\alpha_{xx}=b\delta(y)\sum_n\delta(z-nd)\approx (b/d)\delta(y)$.  In this
superposition of many dislocation
lines, stacked along the $z$ axis, we implicitly took the limit
in which the boundary at large $\vert z\vert$ goes to infinity first:
It is the $q_z=0$ term in the Poisson summation formula
which puts the Dirac $\delta$-function in the expression for $\grad\cross\dnb$
and
not into $\dz\ts$.

Upon inserting these results into \efreeatlast , using our earlier results
for screw dislocations, and neglecting
the interactions between the lines, we find that the threshold chiral coupling
above
which screw dislocations will penetrate is
$\gamma_c=f_{\rm screw}/b$ where $f_{\rm screw}$ includes both elastic
and core energies.   For $\gamma > \gamma_c$ it becomes
energetically favorable for screw dislocations to flood into the crystal and
form
grain boundaries until their density is limited by the repulsive interactions
between screw dislocations.  \ffour\ shows a schematic
phase diagram for the polymer crystal.  The transition to a tilt grain boundary
phase occurs along the $\gamma$ axis when $\gamma'$ is small.
In the following section we will
calculate $\gamma'_c$, the critical chiral coupling for the bond-order
chirality.

The tilt grain boundary texture is also suggested by a continuum elastic
``Debye-H\"uckel''
approach similar to that used in \NT\ and \MN .  This continuum theory requires
length
scales not only large compared to the lattice constants, but also large
compared to
the dislocation spacing.  We
require that the dislocation density $\alpha_{\mu j}$
lead to displacements with
finite elastic energies.  Suppose there is a configuration which
only depends, say, on the $y$-coordinate.  In this case it is
straightforward
to show, using \wdoubleu-\curlpsi\ that the only configurations with
non-divergent
elastic energies are those for which $\alpha_{xx}({\bf q})\rightarrow {\rm
const.}$
as ${\bf q}\rightarrow{\bf 0}$
and all other components of $\alpha_{\mu j}$ vanish in the
same limit.
We consider only configurations with no edge dislocations parallel to the
polymers,
as
they cost a logarithmically divergent energy per unit length.  Upon taking
$\alpha_{\nu j}\propto(2\pi)^2\delta(q_x)\delta(q_z)\delta_{\nu x}\delta_{jx}$
we see
that the only non-vanishing component of $w_{\mu j}$ is $w_{zx} =
i\alpha_{xx}/q_y$ which,
upon
substitution into \efreeii\ gives $\free=iK_2q_0q_y w_{zx}\vert_{{\bf q}={\bf
0}} = -\gamma\alpha_{xx}$.  Note that the quadratic contribution to the elastic
energy in \efreeii\ vanishes in this continuum limit.
We now add the core energies of the dislocations, leading to the free energy
$\free_{DH}$ in the continuum
Debye-H\"uckel
approximation:
\eqn\etgbdh{\free_{\rm DH} =
E_{ijkl}\overline{\alpha_{ij}}\,\overline{\alpha_{kl}}
- \gamma{\bar\alpha_{xx}}}
where $E_{ijkl}$ is a positive definite matrix of line energies of
dislocations
lying in the $xy$-plane and $\overline{\alpha_{ij}}$
is the spatially averaged dislocation density
tensor. As discussed in \MN\ the core energies of screw dislocations
have the form
\eqn\eenerg{E_{ijkl}{\overline{\alpha_{ij}}}\,\overline{\alpha_{kl}} =
E_{\rm screw} \left(\overline{\alpha_{ii}}\right)^2
+ {E'}_{\rm screw}\left(\overline{\alpha_{ij}}\right)^2}
where $E_{\rm screw}$ and ${E'}_{\rm screw}$ are core energies for screw
dislocations.  In the case of a single dislocation ${E'}_{\rm screw}=0$.
The minimum of \etgbdh\
occurs when $\overline{\alpha_{xx}} = \gamma/(2E_{\rm screw}+2{E'}_{\rm
screw})$,
and all other components of $\alpha_{ij}$
vanish.  This is precisely the continuum version of the microscopic  tilt
grain boundary
lattice.
In the same approximation, using the relation between $w_{zx}$ and
$\alpha_{xx}$, we
find $\delta n_x \equiv w_{zx} = -[\gamma/(2E_{\rm screw}+2{E'}_{\rm screw})]
\Theta(y)$,
where
$\Theta(y)$ is the
Heaviside step function, so that across the grain
boundary $\delta n_x$ jumps in proportion to the chiral coupling $\gamma$.
Since Burgers vectors are quantized in units of the lattice vectors, our
proposed periodic array of screw dislocations is the closest allowed
approximation to
a uniform density.

\subsec{Structure Function}
A perfect hexagonal close packed polymer crystal aligned
along the $z$-axis will have six primary $\delta$-function Bragg peaks
in the $q_x$-$q_y$ plane
at a radius  $4\pi/(\sqrt{3}a_0)$, where $a_0$ is the lattice constant
of the polymer crystal.

A tilt grain boundary, being made of parallel screw dislocations, must also be
parallel to
an allowed Burgers vector.  As a result, if two crystalline regions are joined
together by a tilt grain boundary, they will be rotated with respect to
each other around a common reciprocal lattice vector, which we take to be ${\bf
G}
=(4\pi/\sqrt{3}a_0)\left[0,1,0\right]$.  Thus, in Fourier space,
the six spots will rotate around an axis that passes through one pair of
diametric
points on the original hexagon.  If the illumination volume in a diffraction
experiment contains a number of grain boundaries, this
rotation
will continue, laying out the Bragg peaks along two circles lying in the
$q_x$-$q_z$ plane of radius $2\pi/a_0$ at
$q_y=\pm 2\pi/(\sqrt{3}a_0)$.  If the angle of rotation is rational there will
be discrete Bragg spots around the circle, while if the rotation angle is an
irrational
fraction of $2\pi$ the spots will form a continuous circle.  \ffive\ shows
a hypothetical structure function for a TGB state with pitch axis parallel to
$\hat y$
with a rotation angle of $2\pi/7$. A perfectly periodic array of twist grain
boundaries with spacing $d'$ will produce a finely spaced set of additional
peaks
along the $q_y$ axis centered on $4\pi/\sqrt{3}a_0$
at intervals of $2\pi/d'$ (not shown in \ffive).
If the tilt grain boundaries are more irregularly
spaced they will nevertheless limit the range of translational correlations
along $\bf\hat y$ and all
Bragg peaks will be broadened in
the
$q_y$ direction with a width on the order of $2\pi/d'$, where $d'$ is a
translational
correlation length.

We expect that due to long range interactions between the grain boundaries the
angles of rotation will lock in at rational fractions of $2\pi$, in analogy
with the Renn-Lubensky twist grain boundary phase \RLRAT .  While a detailed
calculation of $d$ and $d'$,
in terms of the Landau parameters is difficult, we can estimate their sizes by
assuming that repulsive interactions between screw dislocations lead to
$d\approx d'$.  If the Burgers vector has length $b=a_0$,
the angle of rotation across each grain boundary is
$\phi=\tan^{-1}(a_0/d)$,
and thus the pitch is $P=2\pi d'/\phi$.   Taking $d\approx d'\gg a_0$ we have
$P\approx 2\pi dd'/a_0$.  Upon taking a typical cholesteric pitch to be on the
order
of
$P=5000\angstrom$ and the interpolymer spacing to be $a_0=50\angstrom$, we find
that
$d\approx d' \approx 200\angstrom$ and $\phi\approx 14^{\circ}$.

\newsec{The Moir\'e Phase: Continuum Elastic Theory}

\subsec{Strains, Energies and Displacements}
We have seen that a TGB induces a finite jump in the director and only
ripples in the bond-orientational order parameter which integrate to zero.  We
would like to find a dislocation structure which induces a finite jump
in $\ts$, and thus exploits the chiral coupling $\gamma'$.

An attractive possibility is a honeycomb lattice in a constant $z$-plane
composed of screw dislocations on its
links (see \ffour).  Such a structure
generalizes for three-dimensional systems the grain boundaries
discussed in \PP .  To calculate the strain fields we must superpose
the effect of a collection of finite-length screw dislocations with Burgers
vector $\vec b$.
Consider
a collinear
row of segments, pointing along $\hat x$, each of length $d$, separated by
$2d$.
We
can build up a hexagonal lattice by translating, rotating and superposing the
strains or
twists from this distribution of defects.

Consider the dislocation density generated by the row discussed above, namely
$\alpha_{\gamma i}=b\delta_{\gamma
x}\delta_{ix}\delta(y)\delta(z)\sum_n[
\Theta(x-3nd+d/2)-\Theta(x-3nd-d/2)]$.
The only nonzero component is, in Fourier space
\eqn\fouralph{\alpha_{xx}={2b\sin\left({q_xd/2}\right)\over q_x}\sum_m
e^{3iq_xmd} = (4\pi){b\sin\left({q_xd/2}\right)\over
q_x}\sum_n\delta(3dq_x+2\pi
n), }
where we have used the Poisson summation formula.
When we Fourier transform back to real space, this becomes
\eqn\eposal{\alpha_{xx}(x,y,z) = b\delta(y)\delta(z)\sum_n {\sin(\pi n/3)\over
\pi n} e^{i2\pi n x/3d}.}
We neglect the oscillatory contributions, and approximate the sum
by the $n=0$ term,
\eqn\fouralphii{\alpha_{xx}\approx {1\over 3}b\delta(y)\delta(z).}
Thus the broken line of screw dislocation segments is equivalent at long
wavelengths to
a solid line with $1/3$ the Burgers vector if we are at distances
larger than $d$.  In this approximation, we can replace the
honeycomb lattice with a triangular lattice made by extending all the
edges to meet at the center of each hexagon, provided we divide the
superposition by
$3$ to
restore the correct dislocation density.  Proceeding as in the TGB case, we
have
\eqn\bbcurl{\eqalign{&[\grad\cross\dnb]_{\hbox{moir\'e}} \cr &\;=
-{b\over 6\Delta}\sqrt{\pi\over
\lambda \Delta}
\sum_m \sqrt{\vert m\vert}e^{-\sqrt{\pi m/\lambda \Delta}\vert z\vert}
\sin\left[\sqrt{\pi m\over\lambda \Delta}\vert z\vert + {\pi\over 4}\right]
\sum_{j=1}^3e^{2\pi im{\vec \eta_j\dot {\vec r}_\perp}/\Delta}\cr
&\;\approx -{b\over 3\Delta}\sqrt{\pi\over \lambda \Delta}e^{-\sqrt{\pi/\lambda
\Delta}\vert
z\vert}
\sin\left[\sqrt{\pi\over\lambda \Delta}\vert z\vert + {\pi\over
4}\right]\sum_j\cos[2\pi\vec
\eta_j\dot \vec
r_\perp/\Delta]\cr}}
with $\vec \eta_j = \hat z\times \vec e_j$, where $\vec e_j$
are the unit lattice vectors of the triangular lattice of polymers
and $\Delta=d\sqrt{3}/2$
is
the spacing between parallel Bragg planes in that lattice.

Now, $\dz\ts$ has the $\delta$-function:
\eqn\hcbdzts{[\dz\ts]_{\hbox{moir\'e}} \approx -{b\over
6\Delta}\sum_{j=1}^3\delta(z)+{1\over
2}[\grad\cross\dnb]_{\hbox{moir\'e}}}
To confirm \ior , note that for a honeycomb lattice, $\alpha_{xx}=\alpha_{yy}
={b\delta(z)/(d\sqrt{3})}$, so
$\Tr[\alpha]=2b\delta(z)/d\sqrt{3}=b\delta(z)/\Delta$.

The honeycomb lattice allows us to exploit the $\gamma'$ term in \efreeatlast .
The integral of $\dz\ts$ over
space from a stack of honeycomb lattices separated by a distance $d'$ along the
$z$-axis is $-b/(2\Delta d')$ per unit volume.
The total length per unit volume of honeycomb is $2d/d^2\sqrt{3}d'$.
Thus, assuming $d,d'\gg a_0$ so that interactions among
dislocations are negligible, $\gamma'_c = 2f_{\rm screw}/b$.  The region in
which we
expect
the moir\'e state is shown in \ffour .  In \fseven\ we show a set of polymer
trajectories and the dislocations leading to them.
The moir\'e state in \fseven\ contains an inner region of polymers
which consist of a single polymer at the center
of a bundle of six polymers twisting around it.
This texture has double
twist in the nematic field
as found
in the low-chirality limit of
blue phases of cholesteric liquid crystals \refs{\SWM,\KLEMi}.
The moir\'e state takes advantage of both double twist energies and the
new chiral coupling $\gamma'$.  The bond order field $\ts$ and the nematic
director
$\bf n$ are linked to each other geometrically -- since $\ts$ is measured {\sl
around}
$\bf n$, when the director is not uniform we must carefully define $\ts$.  This
leads
to the same sort of considerations which occur in He$^3-A$:  the wavefunction
is defined in the plane perpendicular to the nuclear spin axis.  The
appropriate
covariant derivative, which take the nematic ``curvature'' into account, is
\eqn\emh{D_\mu\ts = \partial_\mu\ts - \Omega_\mu}
where $\bold{\Omega}$ is the connection and its curl is unambiguously given by
the Mermin-Ho relation:
\eqn\emhh{\left[\curl\bold{\Omega}\right]_{\mu} = {1\over
2}\epsilon_{\mu\nu\rho}\epsilon_{
\alpha\beta\gamma} n_\alpha\partial_\nu n_\beta\partial_\rho n_\gamma}
The geometrical connection between the moir\'e state, the Mermin-Ho relation
and
the nearest-neighbor packing of polymers is discussed in \KNT .

The honeycomb network of screw dislocations described above can also be
understood in terms
of a continuum approach where one looks
for dislocation densities which lead to strains which
only depend on the $z$-coordinate.  As in the TGB phase we neglect
edge dislocations parallel to the polymers and set $\alpha_{zj}=0$.  Upon
assuming
$\alpha_{kj}$ proportional to $(2\pi)^2\delta(q_x)\delta(q_y)$, we find that
$w_{zj}=0$ and
\eqn\ewthing{
w_{kj} = {i\over q_z}
\left[\matrix{\alpha_{yx}&\alpha_{yy}\cr-\alpha_{xx}&-\alpha_{xy}\cr}\right],
\qquad\matrix{
k=x,y\cr j=x,y\cr}}
Upon substituting this result into \efreeii , we find the Fourier transformed
free energy
\eqn\emdh{\eqalign{\free_{\rm DH} &= {\lambda\over
2q_z^2}\left[\alpha_{yx}-\alpha_{xy}\right]^2
+ {\mu\over q_z^2}\left[\alpha_{xy}^2+\alpha_{yx}^2
+\half\left(\alpha_{yy}-\alpha_{xx}\right)^2\right]
+ E_{ijkl}\alpha_{ij}\alpha_{kl}\cr
&\quad - {\gamma'\over 2}\left[\alpha_{yy}+\alpha_{xx}\right]\vert_{{\bf
q}={\bf 0}}\cr}}
In order to eliminate the terms that diverge as $q_z\rightarrow 0$,
we must
take $\alpha_{yx}=\alpha_{xy}$ ({\sl i.e.} no edge dislocations) and
$\alpha_{xy}=\alpha_{yx}
=\alpha_{xx}-\alpha_{yy}=0$.  Thus $\alpha_{xx}=\alpha_{yy}$ are the only
non-zero
components of $\alpha_{kj}$.  The minimum of \emdh\ occurs at
$\overline{\alpha_{xx}}=\overline{\alpha_{yy}}
= \gamma'/(8E_{\rm screw}+4E'_{\rm screw})$ where we have
used \eenerg .  Moreover, $\ts = \half
(w_{xy}-w_{yx}) =
-[\gamma'/(8E_{\rm screw}+4E'{\rm screw})]\Theta(z)$ and so the bond order
jumps
discontinuously as we go across
the
helical grain boundary by an amount proportional to the imposed chirality.
As in the TGB phase the honeycomb array of screw dislocations is the closest
approximation
to a uniform sheet distribution of screw dislocations in terms of quantized
defects with discrete Burgers vectors.

\subsec{Structure Function}

The helical
grain boundary will rotate the six primary Bragg spots of the structure
function
of a perfect crystal in
Fourier space.
In this case all six spots will be swept around a ring in the $q_x$-$q_y$ plane
of
radius $\vert\vec q_\perp\vert = 4\pi/(\sqrt{3}a_0)$.
If the rotation angle is a rational fraction of $2\pi$ there will be a discrete
set
of spots, while if the angle is an irrational fraction of $2\pi$ the spots will
form
a Bragg ring.  While we might expect that interactions between grain boundaries
would favor rational lock-in angles, we shall see that the geometry of an
isolated helical
grain boundary favors an irrational angle.  In \fsix\ we show a schematic
structure function for the irrational angle $\phi_3 =
2\tan^{-1}(\sqrt{3}/21)=9.4^{\circ}$ obtained by superimposing the structure
functions
of
$10$ crystalline regions ({\sl i.e.} $9$ grain boundaries).  An infinite
periodic array (spacing $d'$) of twist grain boundaries would lead to a finely
spaced set of Bragg peaks along the $q_z$ axis, at positions $q_z = 2\pi n/d'$.
 If the
twist boundaries are spaced more randomly, the translational correlations along
$\bf\hat z$ will be still limited to a range of order $d'$.  Reflecting this
fact,
the Bragg spots will be broadened
along the $q_z$ axis, with a width $\approx 2\pi/d'$, as indicated in \fsix .
When many incommensurate twist boundaries are included in an illumination
volume, the
diffraction rods in \fsix\ will merge into a continuous ring, broad along
$q_z$,
but with a very narrow radial
width.  We have verified this numerically by calculating the scattering from a
perfect moir\'e state composed of $24$ grain boundaries.  This Fourier
transform
is axially symmetric about $q_z$ and in \fnumer\ we show the structure in the
$q_y$-$q_z$
plane.

{}From equation \hcbdzts\ we see that across a helical grain boundary the
angles
changes
by $\phi \approx a_0/2\Delta$ (with $b=a_0$ the minimum Burgers vector).
As in the TGB case, an exact calculation of $d$ and $d'$ would require a
detailed accounting
of elastic interactions between screw dislocations.
We can again estimate their size by taking $d\approx d'\gg a_0$.
In this case the pitch is $P= 2\pi\sqrt{3} dd'/a_0$.  If a typical pitch
is $P=5000\angstrom$ and $a_0=50\angstrom$ then we find that $d\approx
d'=150\angstrom$
and $\phi\approx 11^{\circ}$.

\newsec{The Moir\'e Phase: Microscopic Model}

\subsec{Iterated Moir\'e Maps}

When one rotates one lattice with respect to another around a common point, it
is
known that at certain angles there will be a non-zero density of points
coincident
to both lattices \BOL .
By choosing the rotation angle across the
honeycomb dislocation network to produce a high density of coincidence lattice
sites \BOL\
we produce
especially low strain energies across the boundary. The superposition of
triangular polymer
lattices below and above the boundary forms a moir\'e pattern.
\fnine\ shows a sequence of four especially simple angles for a triangular
lattice.   Polymers
in
the lower half-space (circles) must be connected to the closest available
polymer in the
upper half-space (crosses) to minimize bending energy.
The map has a discrete translational
symmetry, in the sense that any coincidence site could be a center of rotation.
Note that an identical moir\'e pattern could have been obtained by rotating
about
a point of $2-$ or $3-$fold symmetry of the coincidence lattice.
Especially
simple moir\'e maps arise for the rotation angles
\eqn\elock{
\phi_n = 2\,\tan^{-1}\left[{\sqrt{3}\over 3(2n+1)}\right]
}
$n=1,2,\ldots$.  \fnine\ shows these maps for $n=1,\ldots,4$, with angles
$\phi_1\approx 21.8^{\circ}$, $\phi_2\approx 13.2^{\circ}$,
$\phi_3\approx 9.4^{\circ}$ and $\phi_4\approx 7.3^{\circ}$.

It is shown in Appendix A that all such angles are irrational fractions of
$2\pi$ \KNW\ so that the structure never repeats upon iteration.
Around each coincidence point
there are $n$ concentric rings of helical polymers.  The lattice of coincidence
points is
also a triangular lattice, but with a spacing $a_n=a_0\sqrt{1+3(2n+1)^2}/2$,
where $a_0$ is the original lattice constant.
The geometrical origin of such energetically preferred lock-in angles has no
analogue
in chiral smectics.  The exact choice of lock-in angles and spacing between
moir\'e planes
must be settled by detailed energetic calculations.
For a {\sl square} lattice an especially simple sequence of lock-in angles
is given by
\eqn\elocsq{
\phi_n^{\rm square}
= \tan^{-1}\left[{2n+1\over 2n(n+1)}\right],
}
leading to coincidence lattice spacings
$a_n^{\rm square}=a_0\sqrt{n^2+(n+1)^2}$ \KNW .

A schematic of the helical grain boundary energy $E(\phi)$ as a function of
angle for
$\gamma=\gamma'=0$ is shown in \fexx.  The deep cusps correspond to
the special lock-in angles discussed above.  See \ref\BKS{R.W.~Balluffi,
Y.~Komem and
T.~Schober, Surf. Sci. {\bf 31}, 68 (1992).} for a discussion of similar
phenomena
in conventional crystals.  When $\gamma'\ne 0$ the energy has the form
\eqn\encvc{\tilde E(\phi) = E(\phi) - c\gamma\phi}
where $c$ is a constant.  As $\gamma$ increases, the preferred minimum
eventually jumps from
the origin to one of the preferred lock-in angles.  In the continuum approach
of Section 4,
we estimated a moir\'e twist of $\phi\approx 11^{\circ}$.  The $n=3$ moir\'e
state nicely
approximates this continuum result with $\phi_3\approx 9.4^{\circ}$.  In
conventional
crystals, cusps in the energy landscape near a lock-in angle $\phi_n$ will have
a contribution
from the density of extra dislocations, proportional to $\phi-\phi_n$ as well
as the logarithmic interaction between the dislocations, leading to the form
$\Delta\epsilon\propto\vert\phi-\phi_n\vert\ln\vert\phi-\phi_n\vert$.  In the
columnar
crystal, due to the softer interaction between dislocations, the leading term
in the
energy will behave as $\Delta\epsilon \propto \vert\phi-\phi_n\vert$ which is
still
sharp enough to produce the lock-in angles.

It is interesting to consider the effect of the moir\'e mapping
on nearest neighbor bonds. Consider a polymer crystal at $z=-\infty$.
In the $xy$-plane the polymers sit at the sites of a triangular lattice.  We
may
draw lines connecting nearest neighbors as shown in \ften(a), with
polymers
at the vertices.  After allowing various arrangements of screw dislocations
to pierce the crystal, we can look again at the
polymers in the
$xy$-plane at $z=+\infty$.  Again, the polymers will sit at the sites of a
triangular
lattice,
typically rotated with respect to the original one.
The effect of intervening defects shows up if we undo the rotation but retain
the original nearest neighbor connectivity of the polymers at $z=-\infty$.
\ften(b) shows the effect of a single screw dislocation, which has
``sheared'' the polymers on either side with respect to the other.  Note that
each vertex still has the coordination number six it had at $z=-\infty$.
As shown in \feleven, when moving a
bond, in any particular plaquette,
we may make one of two types of moves denoted by $\sigma_1$ and $\sigma_2$.
A row of $\sigma_1$ or $\sigma_2$ moves in a triangular lattice represents the
effect of a left or right handed screw dislocation, respectively.
Note that both moves are area
preserving, in
that the area of the parallelogram remains unchanged.  We may represent the
effect
of a plane of defects by making a sequence of moves.  The restriction of
keeping
a
coordination number of six at each vertex amounts to the restriction that
dislocation
lines may never end and that they must meet in triples at $120^{\circ}$.
\ften(c) shows
the effect of a section of a helical grain boundary, near the intersection of
three defect lines.

Upon two iterations of the moir\'e map separating three regions of
polymer crystal, the first coincidence lattice is rotated with respect to
the second coincidence lattice by precisely the angle of rotation $\phi_n$.
Thus the composite coincidence
lattice is the ``coincidence lattice of coincidence lattices'', with lattice
constant $a_n^2/a_0$.   Moir\'e maps iterated $p$ times lead to triangular
composite
coincidence lattices with spacing $a_n(a_n/a_0)^{p-1}$, {\sl i.e.} to
ever sparser lattices of fixed points with intricate fractal
structure in between them.  \ftwelve\ shows the projected polymer
paths for a lock-in angle of a square lattice (with $\phi_1=\tan^{-1}(3/4)
\approx 36.9^{\circ}$) iterated $p=1,\ldots,4$ times.  Note the intricate
fractal
structure which appears after several iterations of the map.

\subsec{Polymer Trajectories and Lyapunov Exponents}

In contrast to
the
TGB state polymers braided by moir\'e maps with  $p\gg 1$ are
highly entangled and wander far from straight line trajectories.
\fseven\ shows the polymer trajectories for the $n=1$ moir\'e map,
of a
triangular lattice iterated nine times. In \fthirteen\ we show forty random
polymer
trajectories after 99 iterations.  Although these polymers are clearly
influenced by
their proximity to a special
center
of rotation, exceedingly complex trajectories are superimposed on their slow
drift around this center.  The center becomes less and less noticeable
for distant polymers.  Using a perfect moir\'e state with $3721$ polymers, we
have calculated the average Fourier transform of the monomer density
$\rho(q_\perp,q_z)$
of a {\sl single} polymer, averaged over all members of the array.
In \numform\ we show contour plots of the intensity
$\langle\,\vert\rho(q_\perp,q_z)\vert^2\,\rangle$
in both the $q_x$-$q_y$ and $q_y$-$q_z$ plane.  Again we see that structure
is approximately axially symmetric about $q_z$.  If a dilute concentration of
deuterated
polymers were introduced into the mix, neutron scattering should produce these
incoherent averages.  In addition, we have studied numerically
the scaling behavior of the polymer size as measured by the components of
moment of inertia tensor in the radial and azimuthal
directions relative to the center
as a function of $N$, the number of iterations.  We expect that
the radius of gyration of the projected polymer, $R$ should be a function of
$N$ and $R_0$, the radius of
the starting point of the polymer trajectory.  We postulate for the radial
radius of
gyration that
\eqn\escli{{R_r(N,R_0)\over R_0}= N^{x_r}\, A_r\!\left({N^{y_r}\over
R_0}\right).}
For large $R_0$ or small $N$, we find that the radial $R_r$ and azimuthal
$R_\phi$
radii of gyration do not depend on $R_0$ and both scale as $N^{\half}$, {\sl
i.e.} as a
simple random walk in the $xy$-plane.  Thus
as $z\rightarrow 0$, $A_r(z)\sim z$ and $x_r+y_r=\half$.
Similarly, the azimuthal radius of gyration scales as
\eqn\escliii{{R_\phi(N,R_0)\over R_0} = N^{x_\phi}\,
A_\phi\!\left({N^{y_\phi}\over R_0}\right),}
and again we find that $x_\phi+y_\phi=\half$ and
$A_\phi(z)\sim z$ as $z\rightarrow 0$. These observations suggest
the one parameter scaling form
\eqn\escali{{R^2_i(N,R_0)\over R_0^{\eta_i}} = \free_i\left({N\over
R_0^{\eta_i}}\right)}
where $\eta_i=1/y_i$, $i=r,\phi$, and the scaling functions $\free_i(z)\sim z$
as
$z\rightarrow 0$.  \ffourteen\ suggests that, superimposed on the small
$N$ ``random walk'' behavior is a slow drift
around the rotation center in the azimuthal direction.  This drift represents
the effect of the distant exceptional coincidence site. As the polymers
spread around the circle at radius $R_0$ they appear to stay confined to an
annular
region of width $\sim R_0$.
For simplicity we take $\eta=\eta_r=\eta_\phi$.  We choose
$\eta$ so that the points of crossover to ballistic (for $R_\phi$) and constant
(for $R_r$) behavior collapse.
We find $\eta=1$ gives the best collapse.  In \ffourteen\ we show the radius of
gyration data for $1364$ polymer trajectories around a supercoincidence site of
$99$ $n=1$ moir\'e maps.  We have binned the data into $30$ equal width annuli
to
and assigned an average $R_0$ to each bin.  The figure clearly shows that
$x_i+y_i=\half$.  After the crossover points $R_r(N,R_0)\rightarrow {\rm
constant}$ and
$R_\phi(N<R_0)\sim N$, as expected.  In the radial case, there is a second
crossover
point in $R_r(N,R_0)$, presumably where the polymers have made a complete
circle around
the supercoincidence site.  This second crossover is evident in the plot only
for small values of $R_0$.

Not only do the polymers spread from their starting point, but they also
diverge
from their neighbors.  \ffifteen\ shows seven polymers which start as nearest
neighbors.
After $99$ iterations they separate from each other, as shown by their
trajectories.  Consider a pseudo-dynamics under which the $z$-axis becomes
time and we consider the trajectories of the polymers as the trajectories of
two-dimensional
particles.  We can look at the average square distance between pairs of
neighboring
polymers and consider the scaling of the separation as a function of $N$, in
the spirit of
measuring a Lyapunov exponent in a real dynamical system.  We have
performed this numerical experiment with the same set of $1364$ polymers used
to
study the radius of gyration.  We calculate
$\Delta=\sqrt{\langle\,\vert\delta \vec r_i - \delta
\vec r_j\vert^2\,\rangle}\equiv\sqrt{\Delta_r^2+\Delta_\phi^2}$, where $\delta
\vec r_i$ is the displacement of
polymer $i$ from its starting point, and where the average is over all $i$ and
$j$
which are nearest neighbors.  Again, we postulate scaling forms for the root
mean square
radial and azimuthal separations $\Delta_r(N,R_0)$ and $\Delta_\phi(N,R_0)$:
\eqn\escld{{\Delta_i(N,R_0)\over R_0} = N^{\tilde x_i} \,B_i\!\left({N^{\tilde
y_i}
\over R_0}\right).}
{}From our analysis we find that, again, for small $N$, $\Delta_i(N,R_0)$ does
not depend
on $R_0$.  Thus $B_i(z)\rightarrow z$ as $z\rightarrow 0$ and, according to
our data, $\tilde x_i + \tilde y_i=0.8$.  Thus, we will have the scaling form
\eqn\esca{{\Delta_i(N,R_0)\over R_0^{0.8\nu_i}} = \tilde b_i\left({N\over
R_0^{\nu_i}}\right)}
where $\nu_i=1/\tilde y_i$, $i=r,\phi$.
We find the best collapse of data for $\nu_r=\nu_\phi=0.3$.
The collapsed data is shown in \fsixteen .  The separation between neighbors,
for
small $N$ is not quite ballistic.  Presumably this is due to the fact that the
polymers stay with their partners for some time before splitting apart as seen
in the
four rightmost paths in \ffifteen .

\newsec{Acknowledgements}

It is a pleasure to acknowledge stimulating conversations with P.C.~Hohenberg,
T.C.~Lubensky, J.F.~Marko, R.B.~Meyer, F.~Spaepen,
P.L.~Taylor, E.L.~Thomas,
and J.~Toner.  RDK acknowledges the hospitality of IBM Research Division,
as well as
support by National Science Foundation Grant
No.~PHY92--45317.  DRN acknowledges the hospitality of Brandeis University,
AT\&T Bell
Laboratories, and Exxon Research and Engineering, as well as support from the
Guggenheim
Foundation and the National Science Foundation, through Grant. No. DMR-9417047,
and in part through the Harvard Materials
Research Science and Engineering Center via Grant No. DMR-9400396.

\appendix{A}{Irrationality of Lock-in Angles}

The special lock in angles that we have found for the triangular lattice are
all
irrational fractions of $2\pi$.  We will prove a more general theorem in this
appendix that
applies to any rotation which leads to a coincidence lattice with a finite
lattice
constant.  An alternate, algebraic proof appears in \KNW .

Consider two lattices with the second one
rotated with respect to the first by an angle $\phi$ around
a lattice point in common (so that if $\phi=0$ the lattices coincide).  At
certain
``magic angles'' $\phi_n$ there will form a lattice of points which are on
both lattices.  This is the coincidence lattice.  Take the magic angle
in question to be $\phi^*$.  If the angle is not an angle of
symmetry of the lattice ({\sl e.g.}, $2\pi n/6$ for a triangular lattice) then
the
spacing of the coincidence lattice will be larger that the spacing on the
original
lattice by a factor of $\zeta$.  Now consider adding a third lattice, rotated
again
by $\phi$ with respect to the second lattice ({\sl i.e.} rotated by $2\phi$
from
the first lattice and by $\phi$ from the second lattice).  The first two
lattices create a coincidence lattice and the second two lattices create
a coincidence lattice.  However, the last two lattices may be obtained by
rotating the first two lattices by $\phi^*$.  Thus the coincidence lattice of
the last two will be rotated by $\phi^*$ with respect to the coincidence
lattice
of the first two.  Therefore the coincidence lattice of all three lattices will
be the coincidence lattice of the two coincidence lattices and so the spacing
of the supercoincidence lattice will be a factor of $\zeta^2$ larger than
the spacing of the original lattice.

This argument can now be extended to the coincidence lattice of $n$ lattices.
The first
$(n-1)$ lattices form a coincidence lattice, as do the last $(n-1)$ lattices.
These
two coincidence lattices are rotated by $\phi^*$ from each other since the
last $(n-1)$ lattices can be obtained by rotating each of the first $(n-1)$
lattices by
$\phi^*$.  Thus the coincidence lattice of $n$ lattices is the coincidence
lattice
of two $(n-1)$ coincidence lattices.  This means that every time we add a
another
rotated lattice, the coincidence lattice spacing grows by a factor of $\zeta$
and
so the coincidence lattice of $(n+1)$ lattices is a factor of $\zeta^n$ larger
than the original lattice spacing.

If the angle $\phi^*$ were a rational fraction of $2\pi$, $\phi^* = 2\pi{p\over
q}$, then
after no more than $q$ iterations the lattices would start to repeat.  Thus
upon
adding the next lattice, the supercoincidence lattice would remain the same,
and
thus the lattice spacing would not grow.  This implies that $\zeta^{q-1}=1$
and so
the
angle $\phi^*$ must be an angle of symmetry of the lattice.  This proves that
the lock-in angles are not rational fractions of $2\pi$.

\vfill\eject
\centerline{\hbox{\hfil\vbox{\offinterlineskip
\hrule
\halign{&\vrule#&\strut\quad#\quad\cr
height2pt&\omit&&\omit&&\omit&\cr
&\omit&&Symmetry Under\hfil&&Symmetry Under\hfil&\cr
&Operator\hfil&&Spatial Inversion\hfil&&Nematic Inversion\hfil&\cr
height2pt&\omit&&\omit&&\omit&\cr
\noalign{\hrule}
height2pt&\omit&&\omit&&\omit&\cr
&${\bf v\equiv\nabb\ts}\hfil$\hfil&&\hfil$+$\hfil&&\hfil$-$\hfil&\cr
height2pt&\omit&&\omit&&\omit&\cr
\noalign{\hrule}
height2pt&\omit&&\omit&&\omit&\cr
&${\bf n}$\hfil&&\hfil$-$\hfil&&\hfil$-$\hfil&\cr
height2pt&\omit&&\omit&&\omit&\cr
\noalign{\hrule}
height2pt&\omit&&\omit&&\omit&\cr
&${\bf v}\dot{\bf n}$\hfil&&\hfil$-$\hfil&&\hfil$+$\hfil&\cr
height2pt&\omit&&\omit&&\omit&\cr
\noalign{\hrule}
height2pt&\omit&&\omit&&\omit&\cr
&${\bf v}\dot(\curl{\bf n})$\hfil&&\hfil$+$\hfil&&\hfil$+$\hfil&\cr
height2pt&\omit&&\omit&&\omit&\cr
\noalign{\hrule}
height2pt&\omit&&\omit&&\omit&\cr
&${\bf n}\dot({\bf v}\dot\nabb){\bf n}$\hfil&&\omit&&\omit&\cr
&${\bf v}\dot({\bf n}\dot\nabb){\bf n}$\hfil&&\hfil$-$\hfil&&\hfil$-$\hfil&\cr
&${\bf n}\dot({\bf n}\dot\nabb){\bf v}$\hfil&&\omit&&\omit&\cr
height2pt&\omit&&\omit&&\omit&\cr
\noalign{\hrule}
height2pt&\omit&&\omit&&\omit&\cr
&${\bf n}\dot(\curl{\bf n})$\hfil&&\hfil$-$\hfil&&\hfil$+$\hfil&\cr
height2pt&\omit&&\omit&&\omit&\cr
\noalign{\hrule}
height2pt&\omit&&\omit&&\omit&\cr
&$({\bf v}\dot{\bf n})[{\bf n}\dot(\curl{\bf
n})]$\hfil&&\hfil$+$\hfil&&\hfil$+$\hfil&\cr
height2pt&\omit&&\omit&&\omit&\cr
\noalign{\hrule}
height.36truein width0pt&\omit&width0pt&\omit&width0pt&\omit
&width0pt\cr
}}\hfil}}
\item{Table 1.} Symmetries of a variety of operators.  Those with parity $(-)$
are
chiral.  Only combinations which are invariant under nematic symmetry are
allowed.

\listrefs
\listfigs

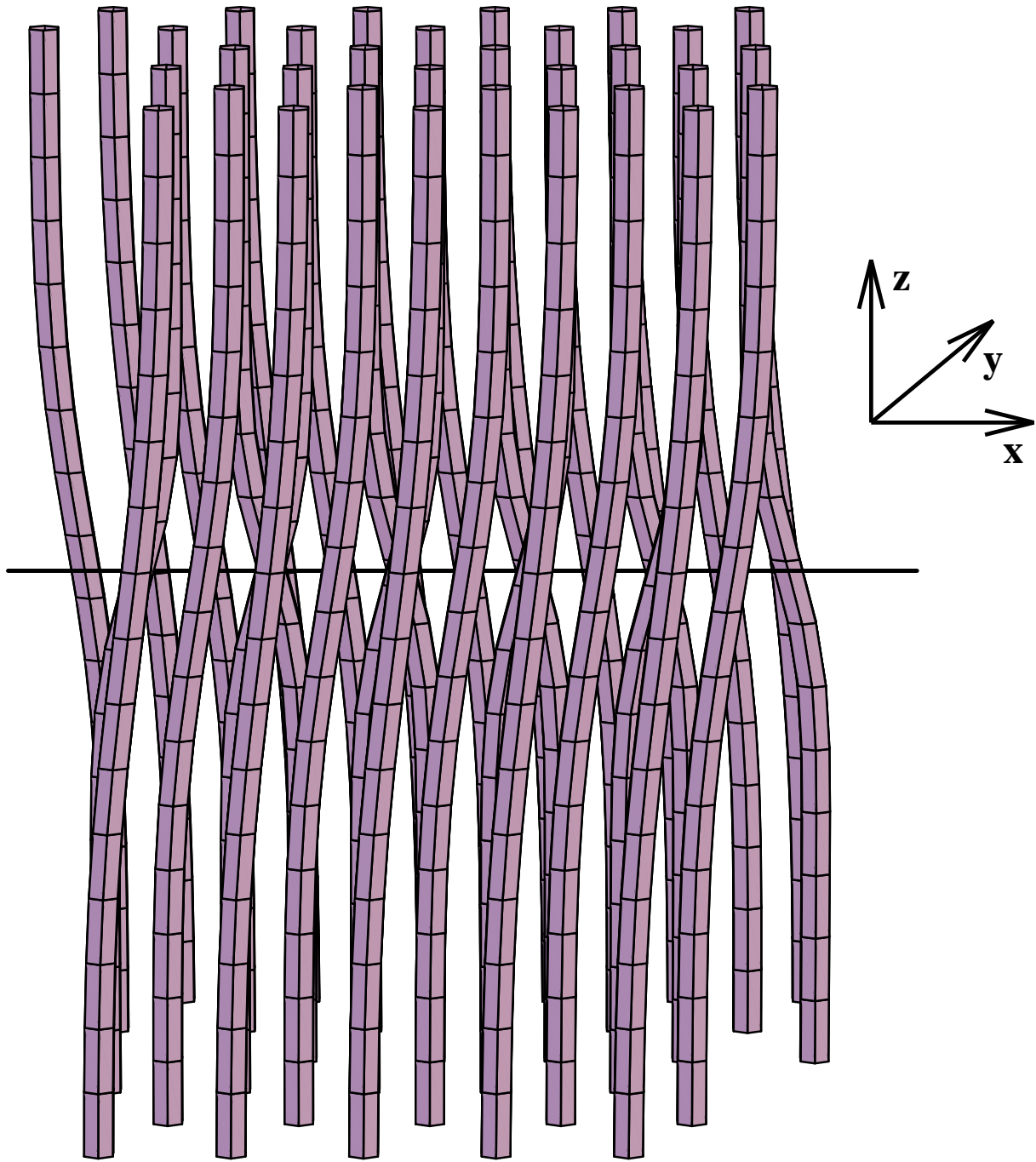
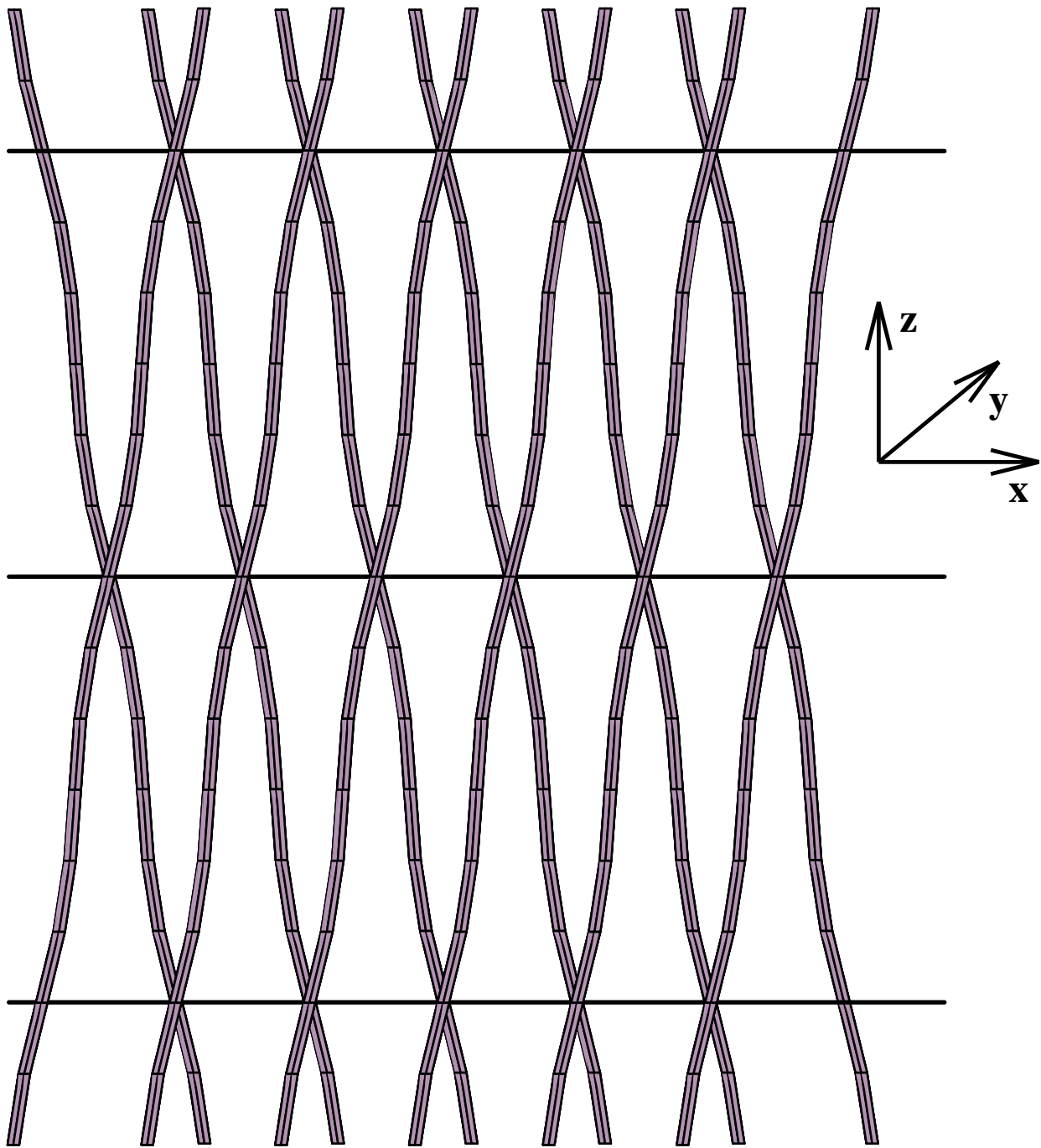
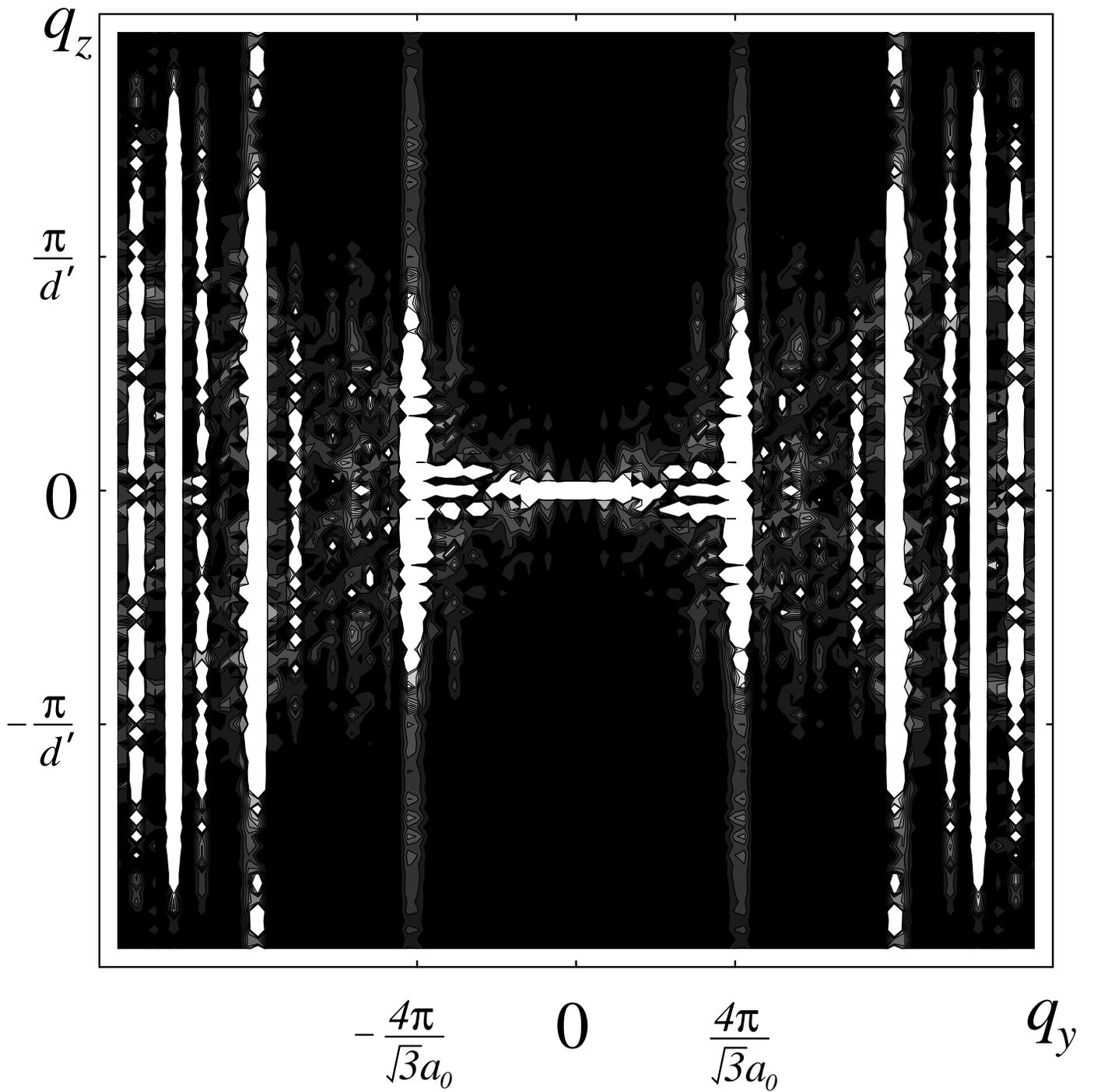
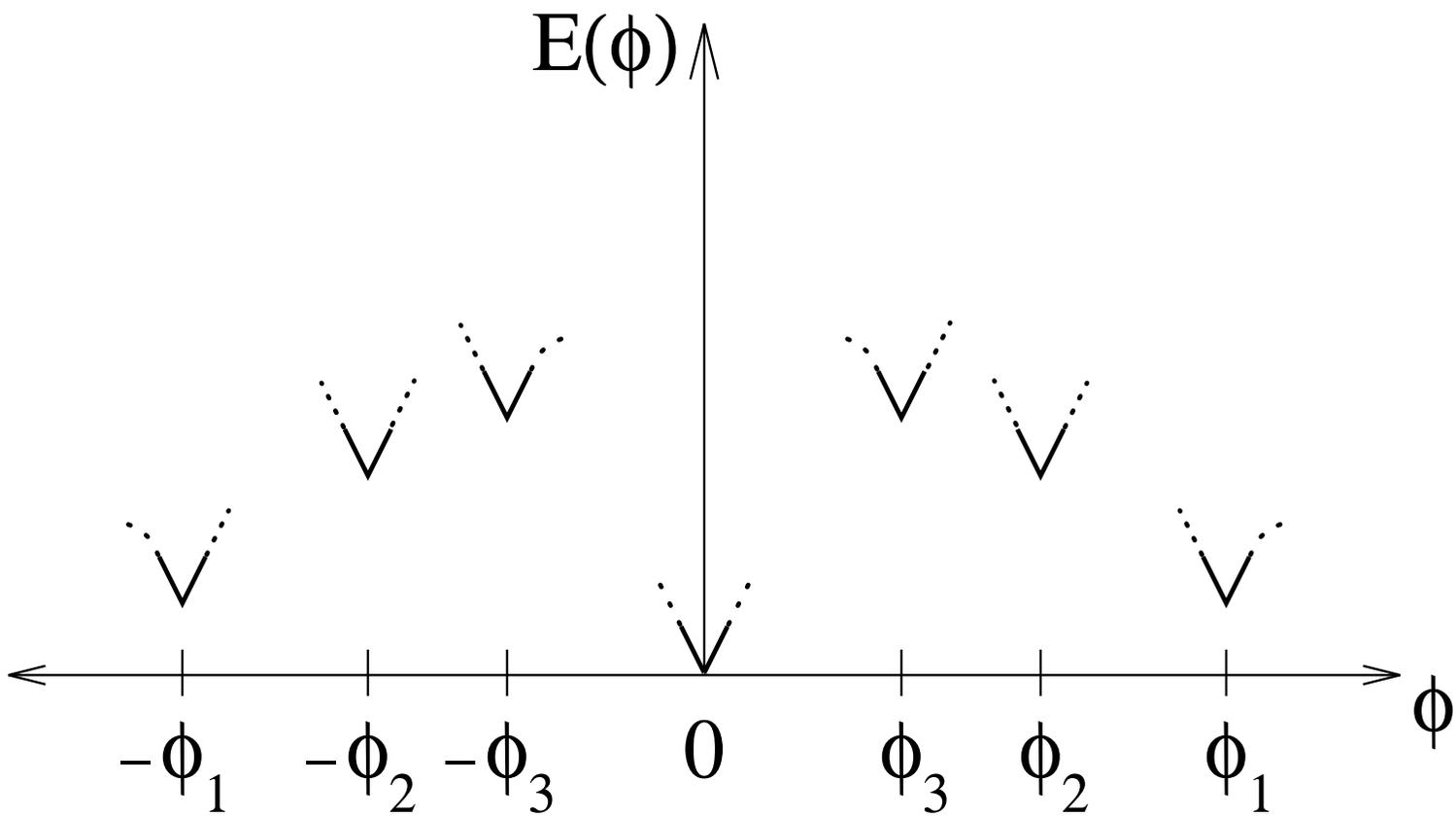
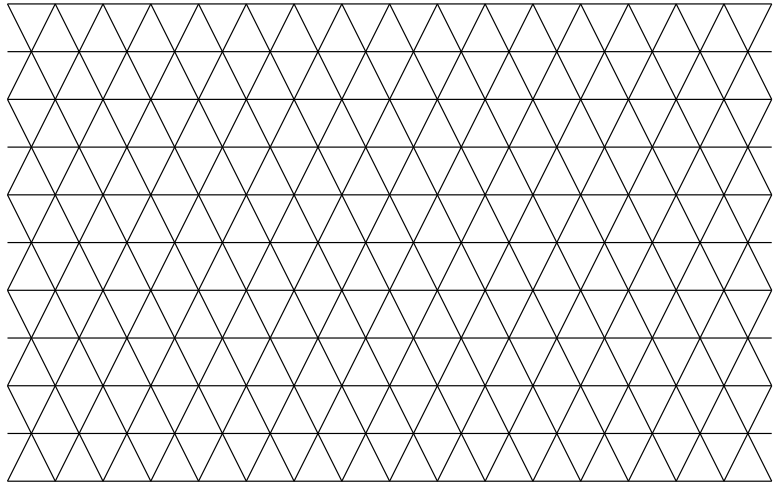
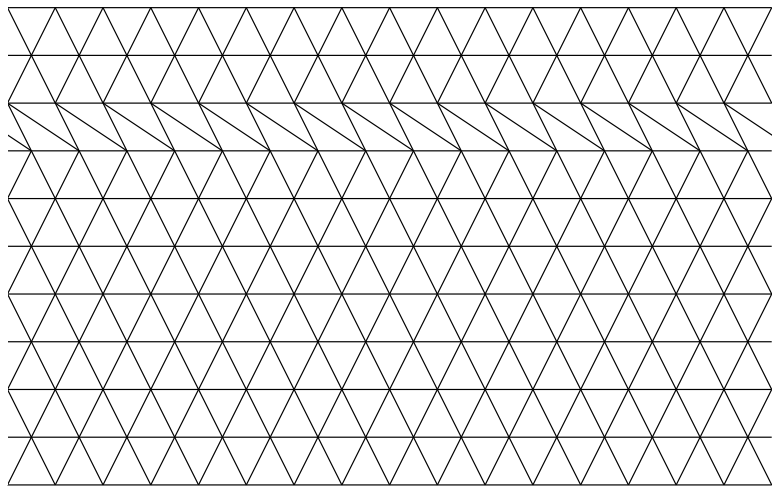
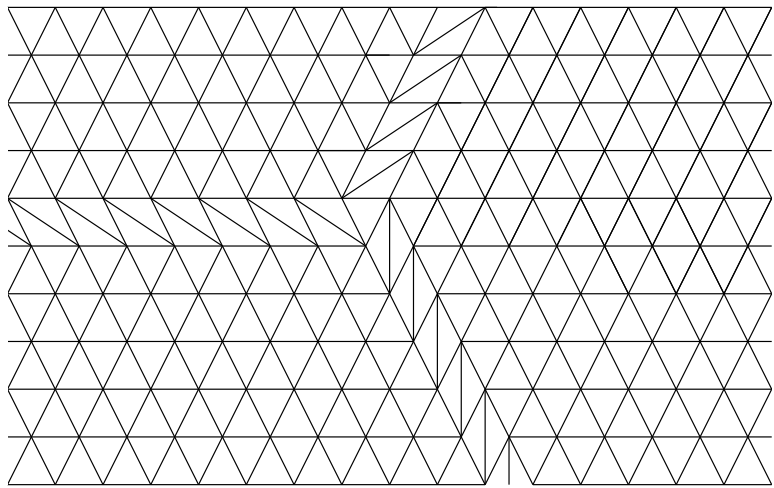
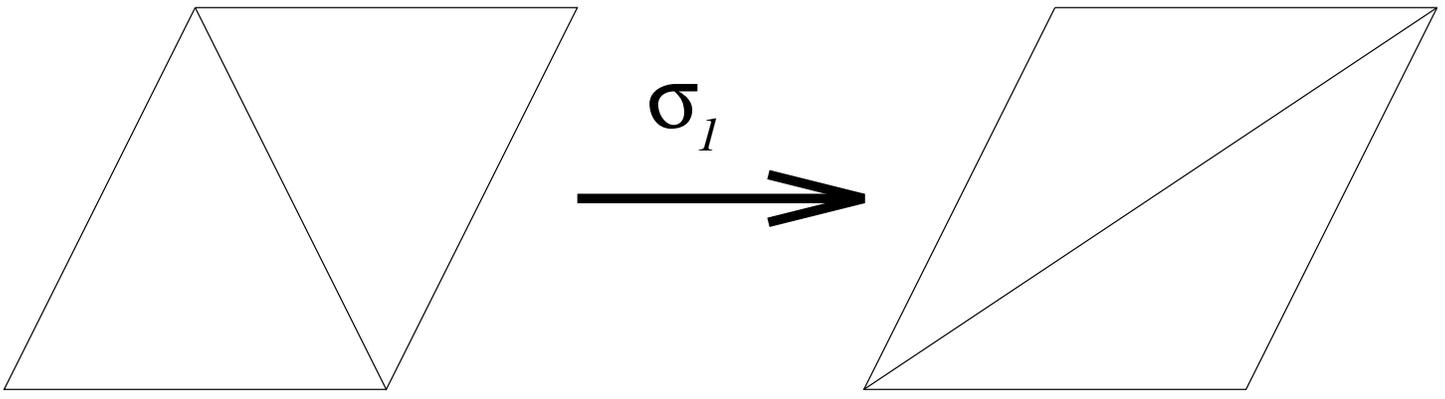
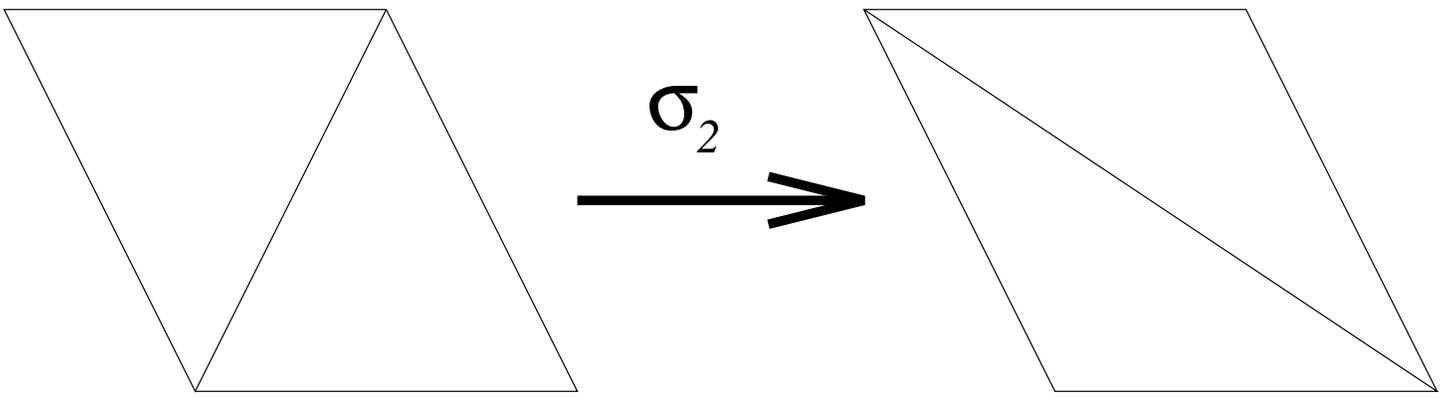
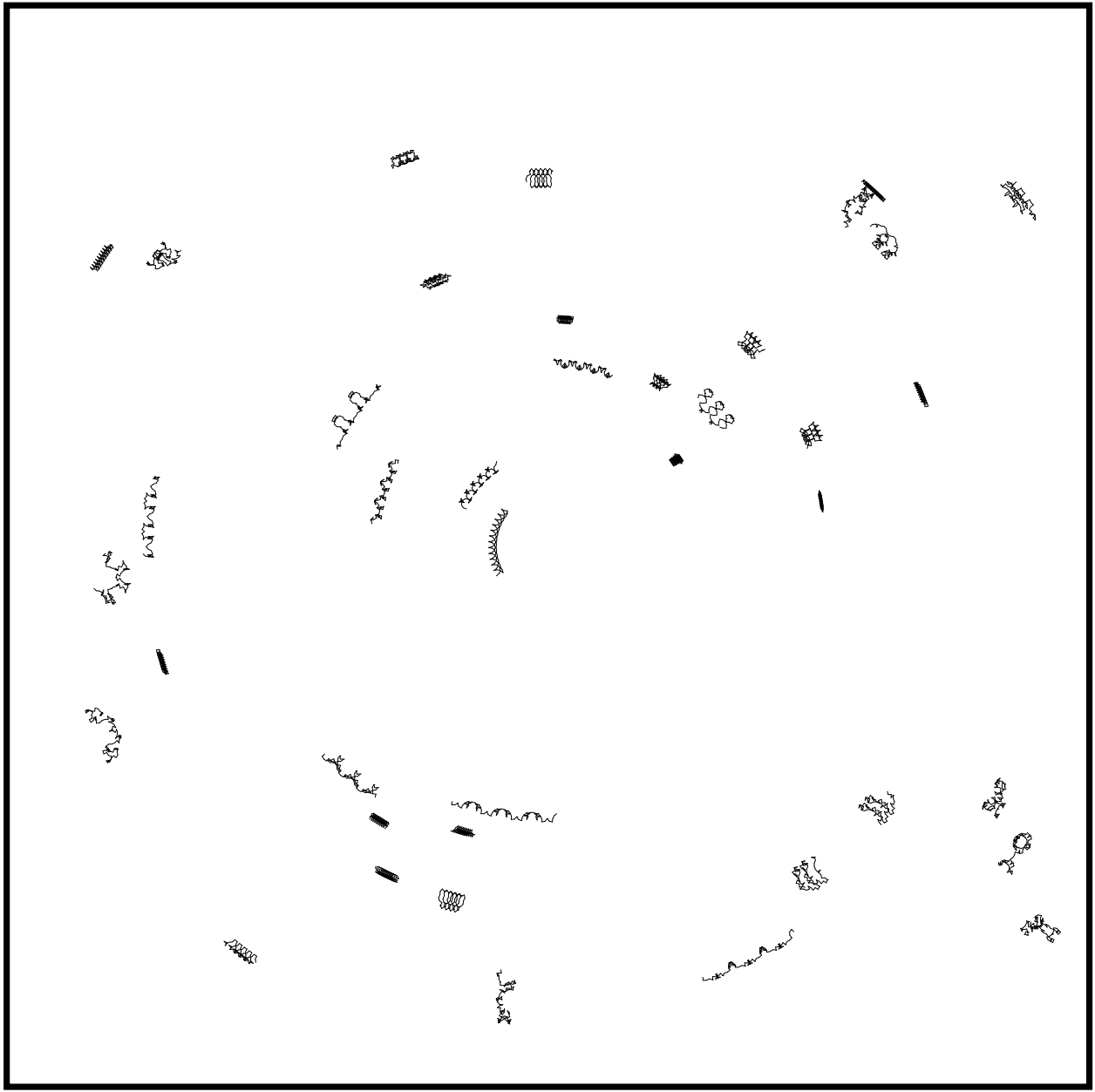
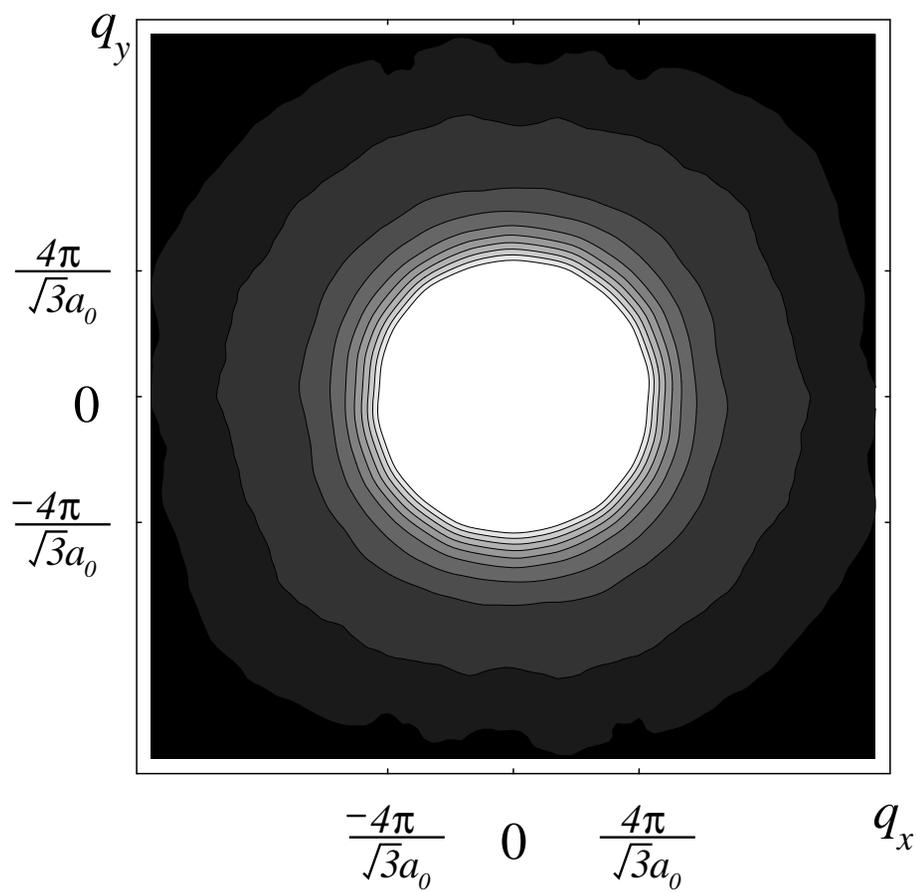
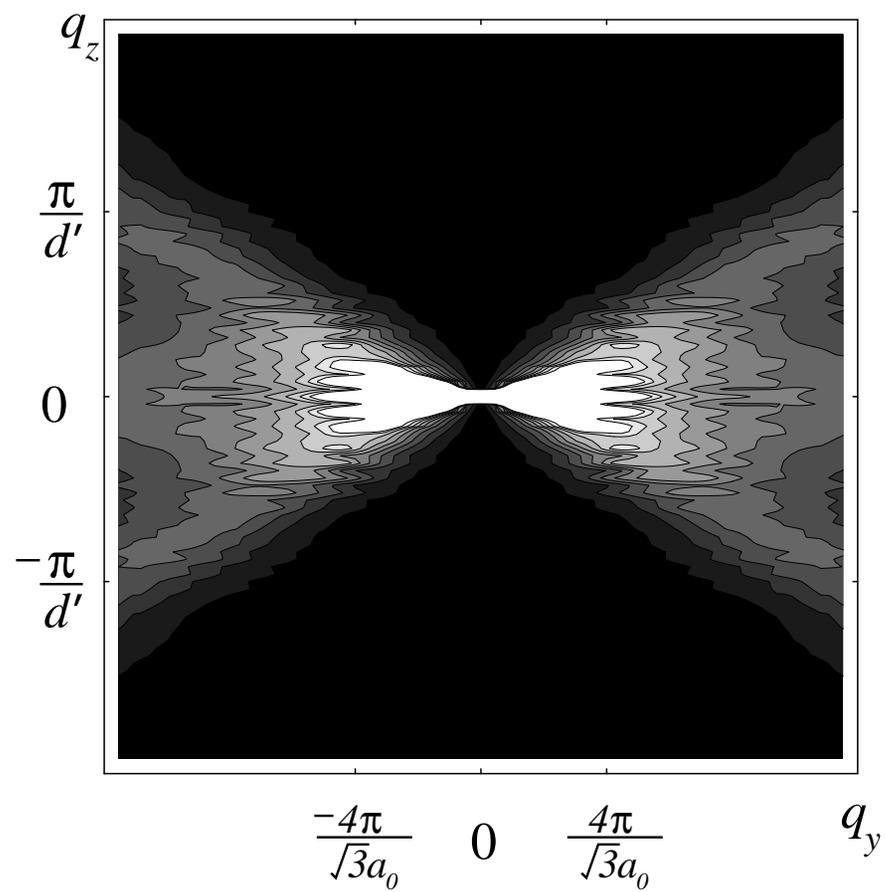
\end